\documentclass[aps,prb,twocolumn,superscriptaddress,floatfix,longbibliography]{revtex4-2}

\usepackage{graphicx,amsmath,amssymb,epstopdf,wasysym}
\usepackage[space]{grffile}
\usepackage{hyperref}
\usepackage[dvipsnames]{xcolor}
\usepackage{braket,amsfonts}
\usepackage{dsfont}
\usepackage{float} 
\usepackage{cancel}
\usepackage{comment}

\usepackage{hyperref}
    \hypersetup
    {   
        unicode=true,           
        pdftoolbar=true,        
        pdfmenubar=true,        
        pdffitwindow=false,     
        pdfstartview={FitH},    
        pdfauthor={Ivan Pasqua, Gabriele Bellomia, Bartomeu Monserrat, Michele Fabrizio, Carlos Mejuto-Zaera},
        pdftitle={TopoGhost},
        pdfkeywords={topological insulators, topological transitions, magnetic ordering, quantum embedding, quasiparticle hamiltonian, ghost Gutzwiller},
        pdfnewwindow=true,      
        colorlinks=true,        
        linkcolor=MidnightBlue,        
        citecolor=Maroon,        
        filecolor=Brown,        
        urlcolor=Maroon          
    }

\newcommand{\be}{\begin{equation}}
\newcommand{\ee}{\end{equation}}

\newcommand{\br}{\mathbf{r}}
\newcommand{\bR}{\mathbf{R}}
\newenvironment{eqs}%
{\begin{equation} \begin{aligned}}%
{\end{aligned} \end{equation} }
\newcommand{\beal}{\begin{eqs}}
\newcommand{\eal}{\end{eqs}}
\newcommand{\eqn}[1]{(\ref{#1})}
\newcommand{\dagga}{{\phantom{\dagger}}}

\newcommand{\fract}[2]{\frac{\displaystyle #1}{\displaystyle #2}}
\newcommand{\bw}{\begin{widetext}}
\newcommand{\ew}{\end{widetext}}
\newcommand{\ep}{\epsilon}

\newcommand{\bealn}{\beal\nonumber}
\newcommand{\Tr}{\mathrm{Tr}}

\newcommand{\coloneq}{\mathrel{\mathop:}\mathrel{\mkern-1.2mu}=}

\begin{document}

\title{Ghost Embedding Bridging Chemistry and One-Body Theories}

\author{Carlos Mejuto-Zaera}
\email{cmejutozaera@irsamc.ups-tlse.fr}
\affiliation{Univ Toulouse, CNRS, Laboratoire de Physique Théorique, Toulouse, France.}

\author{Michele Fabrizio}
\email{fabrizio@sissa.it}
\affiliation{International School for Advanced Studies (SISSA), Via Bonomea 265, I-34136 Trieste, Italy} 

\date{\today}

\begin{abstract}
Phenomenological rules play a central role in the design of chemical reactions and materials with targeted properties.
Typically, these are formulated heuristically in terms of non-interacting orbitals and bands, yet show remarkable accuracy in predicting the complex behavior of intrinsically interacting many-body systems.
While their non-interacting formulation makes them easy to interpret, it potentially hinders the development of new rules for systems governed by strong correlation, such as transition metal-based materials.
In this work, we present a rigorous framework that allows bridging between fully interacting, even strongly correlated, systems and an effective one-body picture in terms of quasiparticles.
Further, we present a computational strategy to efficiently and accurately access the main components of such a description: the embedding approximation of the ghost Gutzwiller Ansatz.
We illustrate the capabilities of this quasiparticle formulation on the Woodward-Hoffmann rules, and apply their reformulated version to toy ``reactions'' which exemplify the main scenarios covered by them.
\end{abstract}

\maketitle

\section{Introduction}

Chemistry and materials science tackle extraordinarily complex many-body systems and, consequently, phenomenological rules often lie at the center of the successful design of new synthesis pathways and devices of tailored opto-electronic properties.
These are often so instrumental that they become a regular part of undergraduate curricula, as is the case with H\"uckel~\cite{Huckel1931a,Huckel1931b,Huckel1932}, Goodenough-Kanamori~\cite{Goodenough1955,KANAMORI1959} or Woodward-Hoffmann rules~\cite{Woodward1965,Longuet-Higgins1965,Zimmerman1966,Woodward1969,Dewar1971,Fukui1982}.
These last ones, for example, concern predicting whether certain types of chemical reactions are likely to be thermally activated or not.
While they can be formulated in different ways, commonly the Woodward-Hoffmann rules are stated in terms of molecular orbital symmetry: In essence, along a symmetry-preserving reaction pathway, if the frontier orbitals involved in the transformation belong to different irreducible representations and cross in the HOMO–LUMO gap, the reaction is not expected to proceed thermally. 
One calls such reactions Woodward-Hoffmann forbidden, while reactions without such a crossing are Woodward-Hoffmann allowed.
This nicely illustrates the nature of such phenomenological rules: they connect the complex chemical reality to a simple, intuitive model in terms of non-interacting molecular orbitals, somehow without compromising their accuracy.
This last point is particularly remarkable when one considers that electron correlation is prevalent in molecules and materials, and crucially becomes dominant along reactions breaking chemical bonds.
This begs the question: Is it possible to derive, or at least convincingly justify, these rules from a fully interacting formalism?
This would not only cement the theoretical underpinnings of already existing rules, but could potentially open a systematic path to discovering new ones tailored to strongly correlated systems, such as transition metal catalysts or quantum materials.

One strategy towards this goal involves justifying these rules using explicitly many-body concepts.
An insightful example concerns recent work by Xie et al.~\cite{Xie2025}, where they observe that the crossing of non-interacting orbitals, central to the original formulation of the Woodward-Hoffmann rules, may directly translate into the crossing of zeros of the one-body Green's function.
While this direction is highly interesting, fully foregoing the non-interacting picture poses two inconvenient challenges: (i) the interpretation, or distilling, of phenomenological rules from the frequency dependent Green's function is more difficult than from an orbital theory, and (ii) accessing the witnesses for the fulfillment/violation of these rules is computationally much more expensive in a fully interacting framework.
Ideally, one would wish for a formulation firmly rooted in the interacting limit, yet somehow leveraging the language of non-interacting orbitals.

This is precisely the philosophy we adopt in this work. 
Starting from a fully interacting perspective, we derive an interpretable, one-body framework which is amenable to formulating phenomenological rules governing correlated electrons and compatible with an efficient computational implementation.
We refer to this as the quasiparticle picture~\cite{mio-2,mio-Mott}.
In essence, the idea is representing the features of strongly correlated electrons using auxiliary non-interacting systems, as one does in Kohn-Sham density functional theory for weakly-correlated materials~\cite{martin2004}.
Within such a quasiparticle language, it is possible to justify and formulate phenomenological rules in terms of molecular orbitals, yet remaining by construction in a fully interacting framework.
We present the theoretical ingredients of this quasiparticle formalism, together with a computational strategy to efficiently and accurately access its main witnesses: the embedding approximation of the ghost Gutzwiller Ansatz~\cite{Lanata2017,guerci2019,frank2021,Mejuto2023a,Lee2023a,Lee2023b,Guerci2023}.
We will exemplify this programme on the Woodward-Hoffmann rules, using two toy ``reactions'' to illustrate how they can be reformulated in the quasiparticle language.
Looking ahead, this strategy has the potential to unveil new phenomenological rules for strongly correlated molecules and materials.

\section{An alternative view on the Woodward-Hoffmann rules}
We start by reexamining the arguments of \cite{Xie2025} from a different perspective, partly following Refs.~\cite{mio-Mott,Andrea-PRB2023,Ivan-SciPost}. We assume a generic basis of single-particle molecular orbitals $\psi_\alpha(\br)$, where $\alpha=1,\dots,2N$ includes the spin. 
Although the basis is complete only for $N\to\infty$, we will work with a finite $N$ henceforth.  The zero-temperature Green's function of the imaginary frequency $i\ep$, $\ep\in[-\infty,\infty]$, is generally a matrix $G(i\ep)=G(-i\ep)^\dagger$ in this basis, with elements 
\beal
G_{\alpha\beta}(i\ep) &= \sum_{n>0}\,\Bigg\{
\fract{\;\bra{0}c^\dagga_\alpha\ket{n}\bra{n}c^\dagger_\beta\ket{0}\;}
{i\ep-E_n+E_0}\\
&\qquad \qquad \qquad \quad + \fract{\;\bra{0}c^\dagger_\beta\ket{n}\bra{n}c^\dagga_\alpha\ket{0}\;}
{i\ep+E_n-E_0}\Bigg\}\,,
\label{G Matsubara}
\eal
where $n\geq 0$ runs over all many-body eigenstates $\ket{n}$ with eigenvalues $E_n$. The ground state is identified by $n=0$ and assumed to be non-degenerate. We emphasize that zero imaginary frequency, $\ep=0$, separates the processes of adding an electron to the ground state ($\ep>0$) from those of removing one ($\ep<0$).\\
We write $G(i\ep)$ in \eqn{G Matsubara} as 
\be
G(i\ep) = G_1(\ep)+i\,G_2(\ep)
= G_1(\ep)-i\,\ep\,\Lambda(\ep)\,,
\label{G1 and G2}
\ee
with hermitian  
$G_1(\ep)=G_1(-\ep)$ and $G_2(\ep)=-G_2(-\ep)$, 
and positive definite $\Lambda(\ep)=\Lambda(-\ep)$.
Upon defining 
\be
\Xi(\ep) \coloneq G(i\ep)\,G(i\ep)^\dagger\,,
\label{Xi}
\ee
which is also positive definite, we can represent $G(i\ep)$ via polar decomposition as 
\be
G(i\ep) = \Xi(\ep)^{1/2}\;U(i\ep)\,,
\label{G-SVD}
\ee
with unitary 
\be
U(i\ep) = \Xi(\ep)^{-1/2}\;G(i\ep) \,.
\label{SVD U}
\ee
We next define the hermitian matrices
\beal
&K_1(\ep) = K_1(-\ep) = -\fract{1}{2}\,\Big(G(i\ep)^{-1} + G(i\ep)^{\dagger\,-1}\Big)\\
&\qquad \qquad =  -G(i\ep)^{-1}\,G_1(\ep)\,
G(i\ep)^{\dagger\,-1}
\,,\\
&K_2(\ep) = -K_2(-\ep) = \fract{1}{2i}\,\Big(G(i\ep)^{-1} - G(i\ep)^{\dagger\,-1}\Big)\\
&\qquad\qquad  = -G(i\ep)^{-1}\,G_2(\ep)\,
G(i\ep)^{\dagger\,-1}  \,,
\label{K1 and K2}
\eal
and the \textit{quasiparticle} residue $Z(\ep)$ \cite{Andrea-PRB2023,Ivan-SciPost} 
\be
Z(\ep) = Z(-\ep)=\ep\,K_2(\ep)^{-1} = 
G(i\ep)^\dagger\;\Lambda(\ep)^{-1}\;G(i\ep)\,,
\label{Z1}
\ee
which is semi positive definite with eigenvalues $\in\![0,1]$. Through \eqn{G-SVD} and \eqn{SVD U}, we can rewrite \eqn{Z1} as 
\be
Z(\ep) =  A(\ep)^\dagger\,A(\ep)\,,
\label{true Z}
\ee
where
\be
A(\ep) = \Lambda(\ep)^{-1/2}\,\Xi(\ep)^{1/2}\, U(i\ep)\,.
\label{A-SVD}
\ee
By definition, 
\beal
&G(i\ep)^{-1}= -K_1(\ep) + i\,K_2(\ep) \\
&\qquad= i\ep\,A(\ep)^{-1}\,A(\ep)^{\dagger\,-1} - K_1(\ep)\\
&\qquad= A(\ep)^{-1}\,\Big\{i\ep - A(\ep)\,K_1(\ep)\,A(\ep)^\dagger\Big\}\,
A(\ep)^{\dagger\,-1}\\
&\qquad\coloneq A(\ep)^{-1}\,\Big\{i\ep - H_*(\ep)\Big\}\,
A(\ep)^{\dagger\,-1}\,,
\label{G vs K}
\eal 
thus
\beal
G(i\ep) &= A(\ep)^\dagger\;\fract{1}{\;i\ep-H_*(\ep)\;}\;
A(\ep) \\
&\coloneq A(\ep)^\dagger\;G_*(i\ep)\;
A(\ep)\;,
\label{G vs G*}
\eal
where $G_*(i\ep)$ and 
\beal
H_*(\ep) &= H_*(\ep)^\dagger= H_*(-\ep)=A(\ep)\,K_1(\ep)\,A(\ep)^\dagger\\
&= -\Lambda(\ep)^{-1/2}\,
\;G_1(\ep)\;\Lambda(\ep)^{-1/2}
\;,
\label{H* SVD}
\eal
are, respectively, the \textit{quasiparticle} Green's function 
and Hamiltonian. We remark that the above expressions of $Z(\ep)$ and 
$H_*(\ep)$ are fully equivalent to those in \cite{Andrea-PRB2023} and \cite{Ivan-SciPost}, even though they are here derived without any reference to the self-energy. This alternative derivation may be more convenient when calculating $G(i\epsilon)$ directly is possible within some controlled approximation. \\
The \textit{quasiparticle} Hamiltonian \eqn{H* SVD} depends parametrically on the frequency $\ep$. Specifically, for $\epsilon \to \pm \infty$, $H_*(\epsilon) \to H_\text{HF}$, where $H_\text{HF}$ is the Hartree-Fock Hamiltonian. However, there is a caveat to consider: the expectation values that define $H_\text{HF}$ are evaluated on the true ground state, not the Hartree-Fock one. Consequently, the number of negative eigenvalues of $H_\text{HF}$ is generally not equal to the number of electrons.  \\

\noindent
Employing the exact representation \eqn{G vs G*} of $G(i\ep)$, let us now discuss the topological invariant studied in \cite{Xie2025}. We first notice that, through \eqn{G Matsubara}, 
\beal
G_{1,\alpha\alpha}(\ep) &= \sum_{n>0}\,
\Bigg\{
-R^+_{\alpha,n}\;
\fract{\;E_n-E_0\;}
{\;\ep^2+(E_n-E_0)^2\;}\\
&\qquad \qquad 
+R^-_{\alpha,n}\;
\fract{\;E_n-E_0\;}
{\;\ep^2+(E_n-E_0)^2\;}\Bigg\}\\
&\coloneq -G^+_{1,\alpha\alpha}(\ep) + G^-_{1,\alpha\alpha}(\ep)\,,
\label{G1 Matsubara}
\eal
where 
\beal
R^+_{\alpha,n} &\coloneq \big|\bra{n}c^\dagger_\alpha\ket{0}\big|^2\,,&
R^-_{\alpha,n} &\coloneq \big|\bra{n}c^\dagga_\alpha\ket{0}\big|^2\,.
\label{weights}
\eal
It follows that, since $G_2(\ep)$ is odd,  
\beal
&\int_{-\infty}^\infty 
\fract{d\ep}{2\pi}\,\Tr\big(G(i\ep)\big)
= \int_{-\infty}^\infty 
\fract{d\ep}{2\pi}\,\Tr\big(G_1(\ep)\big)\\
&\qquad= \fract{1}{2}\,\sum_{\alpha=1}^{2N}\,\sum_{n>0}\,
\Big(-R^+_{\alpha,n} + R^-_{\alpha,n}\Big)\\
&\qquad = \fract{1}{2}\,\sum_{\alpha=1}^{2N}\,\big(
-\bra{0}c^\dagga_\alpha\,c^\dagger_\alpha\ket{0}
+\bra{0}c^\dagger_\alpha\,c^\dagga_\alpha\ket{0}\big)\\
&= \mathcal{N}_{els}-N\,,
\label{sum rule}
\eal
where $\mathcal{N}_{els}$ is the number of electrons, which is even since we assume a non-degenerate ground state and time-reversal symmetry, hence with no spin polarization. Alternatively, we can write \eqn{sum rule} through \eqn{G Matsubara} and \eqn{weights} also as
\be
\int_{-\infty}^\infty 
\fract{d\ep}{2\pi}\,\Tr\big(G(i\ep)\big) = 
\int_{-\infty}^\infty 
\fract{d\ep}{2\pi}\,\fract{\partial \ln\text{det}\,\left(\mathcal{G}(i\ep)^{-1}\right)}{\partial i\ep}
\,,
\label{G Matsubara diagonal}
\ee
using the formal definition 
\bealn
\text{det}\,\left(\mathcal{G}(i\ep)^{-1}\right)
&\coloneq \prod_{n>0}\,
\big(i\ep-E_n+E_0\big)^{\sum_\alpha\! R^+_{\alpha,n}}\\
&\qquad\qquad
\big(i\ep+E_n-E_0\big)^{\sum_\alpha \! R^-_{\alpha,n}}\,,
\eal
which demonstrates that \eqn{G Matsubara diagonal} is a boundary term in frequency.\\
In conclusion, the electron number can be written as   
\beal
\mathcal{N}_{els}&= N +
\int_{-\infty}^\infty 
\fract{d\ep}{2\pi}\,\fract{\partial \ln \text{det}\,\left(\mathcal{G}(i\ep)^{-1}\right)}{\partial i\ep}
\\
& \equiv  N +\int_{-\infty}^\infty \fract{d\ep}{2\pi}\;
\fract{\partial \ln \text{det}\left(G(i\ep)^{-1}\right)}{\partial i\ep}
+ I_L\,,
\label{Luttinger}
\eal
where $I_L$ is defined by comparing the two sides of the equivalence in \eqn{Luttinger} and it is  also a pure boundary term \cite{Jan}. The celebrated Luttinger's theorem \cite{Luttinger} states that $I_L=0$ if the ground state can be accessed perturbatively in the electron-electron interaction. However, Luttinger's theorem is violated and $I_L\not=0$ when perturbation theory breaks down \cite{Jan}, which we suspect is generally the case in molecules where correlations effects are strong. \\ 
With the above definitions and through \eqn{G vs G*}, the topological invariant introduced in \cite{Xie2025} is simply 
\beal
\mathcal{N} &= \mathcal{N}_{els}-I_L\\
&=N-\int_{-\infty}^\infty \fract{d\ep}{2\pi}\;
\fract{\partial \ln \text{det}\big(G(i\ep)\big)}{\partial i\ep}\\
&=N-\int_{-\infty}^\infty \fract{d\ep}{2\pi}\;
\fract{\partial \ln \text{det}\big(G_*(i\ep)\big)}{\partial i\ep}\\
&\qquad - \int_{-\infty}^\infty \fract{d\ep}{2\pi}\;
\fract{\partial \ln \text{det}\big(Z(\ep)\big)}{\partial i\ep}\;.
\label{N-1}
\eal
Since $Z(\ep)=Z(-\ep)$, the derivative in the last term of \eqn{N-1} is odd in $\ep$, and therefore vanishes upon integration. It follows that, 
if $\ep_\ell(\ep)=\ep_\ell(-\ep)$, $\ell=1,\dots,2N$, are the eigenvalues of $H_*(\ep)$ in \eqn{H* SVD}, then 
\beal
\mathcal{N} &=N-\int_{-\infty}^\infty \fract{d\ep}{2\pi}\;
\fract{\partial \ln \text{det}\big(G_*(i\ep)\big)}{\partial i\ep}\\
&= N+\sum_{\ell=1}^{2N}\, \int_{-\infty}^\infty \fract{d\ep}{2\pi}\; 
\fract{\partial\ln\big(i\ep-\ep_\ell(\ep)\big)}{\partial i\ep} \\
&\qquad = \sum_{\ell=1}^{2N}\,\theta\big(-\ep_\ell(0)\big)\,.
\label{N-2}
\eal
Therefore, $\mathcal{N}$ is an integer that coincides with the number of negative eigenvalues of the \textit{quasiparticle} Hamiltonian $H_*(\ep)$ 
at $\ep=0$. Equivalently, $\mathcal{N}$ is the number of \textit{quasiparticles} in the ground state of $H_*\coloneq H_*(\ep=0)$, 
and it is also equal to the number of electrons $\mathcal{N}_{els}$ if Luttinger's theorem holds. \\

\noindent
Following \cite{Xie2025}, let us suppose that the molecule undergoes a transformation from an initial state to a final one. We can imagine that the quasiparticle Hamiltonian depends on a generalized coordinate $\bR$, for example, the atomic coordinates. This coordinate changes along a selected path during the transformation from $\bR_i$ in the initial state to $\bR_f$ in the final state. Correspondingly, the topological invariant \eqn{N-2} becomes a function of $\bR$, 
\be
\mathcal{N}(\bR) = \sum_{\ell=1}^{2N}\,\theta\big(-\ep_\ell(\bR)\big)\,,
\label{N-3}
\ee
where $\ep_\ell(\bR)$ are the eigenvalues of 
\be
H_*(\bR)\coloneq  H_*(\ep=0,\bR)\,.
\label{H* R}
\ee
We emphasize that the \textit{quasiparticle} Hamiltonian must be invariant under the same symmetry group G($\bR$) of the physical Hamiltonian, which contains at least time-reversal symmetry $\mathcal{T}$ besides the identity. It follows that $G_*(i\ep,\bR)$ and $H_*(\ep,\bR)$ for any $\ep$ are block matrices, each block representing a different irreducible representation (irrep) $\rho$ of $\text{G}(\bR)$. Consequently, 
\be
\mathcal{N}(\bR) = \sum_{\rho}\, \mathcal{N}_\rho(\bR)\,,
\label{N-4}
\ee
where 
\be
\mathcal{N}_\rho(\bR) = \sum_\ell\, \theta\big(-\ep_{\rho\ell}(\bR)\big)\,,
\ee
with $\ep_{\rho\ell}(\bR)$ the eigenvalues of $H_*(\bR)$ in \eqn{H* R} corresponding to eigenstates that transform like the irrep $\rho$. We note that each eigenvalue  
has at least a Kramers's degeneracy because of time-reversal symmetry. \\
Since $H_*(\bR)$ is effectively a non-interacting Hamiltonian, it is tempting to generalize the Woodward-Hoffmann rules and assume that the molecular reaction is symmetry-forbidden if along the selected path eigenvalues corresponding to different irreps cross each other at zero energy. 
If $\bR_c$ is the value of $\bR$ at which the crossing occurs, then
\beal
0 &=\text{det}\big(H_*(\bR_c)\big) \\
&= \text{det}\big(Z(\ep=0,\bR_c)\big)\,
\text{det}\big(K_1(\ep=0,\bR_c)\big)\,,
\label{roots}
\eal
where we use equations \eqn{G vs K} and \eqn{H* SVD}. It follows that the eigenvalues that cross zero may either correspond to poles or to zeros of $\text{det}\big(G(i\ep=0,\bR_c)\big)$, which is precisely the conclusion of \cite{Xie2025}. The intriguing aspect of our exact representation \eqn{G vs G*} of the Green’s function is that both seemingly contradictory situations can be described by the single \textit{quasiparticle} Hamiltonian. 
In particular, if the instantaneous many-body ground state remains non-degenerate along the adiabatic transformation $\bR_i\to\bR_f$, precisely the case discussed in \cite{Xie2025}, all the eigenvalues of $H_*(\bR)$ that cross zero energy at $R_c$ must correspond to roots of $\text{det}\big(G(i\ep=0,\bR_c)\big)$. Otherwise, the instantaneous ground state would inevitably be degenerate. \\
Moreover, \eqn{G vs G*} suggests a further generalization of Woodward-Hoffmann rules. If Luttinger's theorem is obeyed along the full reaction-pathway, $\mathcal{N}(\bR)$ must be conserved and equal to $\mathcal{N}_{els}(\bR)$. This implies that the number of eigenvalues of $H_*(\bR)$ that cross zero energy upwards must be equal to the number that cross zero energy downwards. \\
On the contrary, $\mathcal{N}(\bR)$ is no longer conserved if Luttinger’s theorem is violated. In this case, $\mathcal{N}(\bR)$ may well have jumps between different integer values along the reaction pathway, each jump corresponding to eigenvalues of $H_*(\bR)$ crossing zero energy. It is worth noting that these eigenvalues come at least in Kramers’ pairs, implying that $\mathcal{N}(\bR)$ jumps by an even integer. If the 
instantaneous ground state is non-degenerate, the eigenvalues of 
$H_*(\bR)$ that cross zero energy still correspond to roots of 
$\text{det}(G(i\ep=0,\bR))$. Since a zero of $\text{det}(G(i\ep=0,\bR))$ implies a singularity of the self-energy at zero frequency, it seems natural to conjecture that the transformation $\bR_i\to \bR_f$ is again forbidden along the selected path.

\section{Model and Methods}

\subsection{The ghost Gutzwiller Approximation}

The main object we are after is thus the quasiparticle Hamiltonian in Eq.~\eqref{H* R}, for which we need to find reasonably accurate yet accessible approximations to Eq.~\eqref{H* SVD}.
It turns out that this quantity can be accessed directly by the ghost Gutzwiller (gGut) Ansatz~\cite{Lanata2017,guerci2019,frank2021,Mejuto2023a,Lee2023a,Lee2023b,Guerci2023}, a generalization of the Gutzwiller wave function~\cite{Gutzwiller1963,Gutzwiller1965,Bunemann1998,Fabrizio2007,Yao2014,Yao2015,lanata2015,fabrizio2017}.
This is a variational, non-perturbative wave function Ansatz which represents the many-electron system ground state wave function $\ket{\Psi_G}$ in terms of a mean-field (quasiparticle) Slater determinant $\ket{\psi_{qp}}$ and a projector operator $P$ as $\ket{\Psi_G} = P\ket{\psi_{qp}}$.
The Slater determinant, often called quasiparticle wave function, can be defined as the ground state of the \emph{quasiparticle Hamiltonian} $H_{qp}$, which we will identify with $H_*$ in Eq.~\eqref{H* R} for reasons that will become apparent later. 

The parameters defining our Ansatz, i.e. the parameters in $P$ and $H_{qp}$, are to be variationally optimized to minimize the energy.
Within an infinite dimensional limit approximation, this variational optimization can be exactly substituted by a self-consistent embedding of $H_{qp}$ into a local impurity problem~\cite{lanata2015}, in a way quite analogous to embedding approximations like dynamical mean-field theory~\cite{Kotliar1996,Kotliar2006,Kotliar2001b,Arita2007,Shim2007,Takizawa2009,Haule2010,Park2014,Haule2015,Paul2019,Zhu2020,Zhu2021}, density matrix embedding theory~\cite{Knizia2012,Wouters2016,Zhu2020b,Sekaran2021,Sekaran2023}, energy-weighted density matrix embedding theory~\cite{Fertitta2018,Sriluckshmy2021} or self-energy embedding theory~\cite{Kananenka2015,Lan2015,Iskakov2020}.
While we defer a detailed description of the method and its algorithmic implementation to the existing literature~\cite{Mejuto2023a,Mejuto2024,Pasqua2025}, we summarize here the main ingredients relevant for the study of Woodward-Hoffmann rules in molecular reactions, and present a bullet-point summary of the algorithm in the SI.

We assume that the system of interest has exclusively local interactions, such that its Hamiltonian follows

\beal
        H &= H^{hop} + \sum_I H^{loc}_{I}, \\
        H^{hop} &= \sum_{I\neq J}\sum_{\alpha_I\beta_J}\, t_{\alpha_I\beta_J}\, c^\dagger_{\alpha_I}\,c^\dagga_{\beta_J} \\
        H^{loc}_{I} &= \sum_{\alpha_I\beta_I}\, t_{\alpha_I\beta_I}\,c^\dagger_{\alpha_I}\,c^\dagga_{\beta_I} \\
        &+\frac{1}{2}\sum_{\alpha_I\beta_I\,\gamma_I\delta_I}\, U_{\alpha_I\beta_I\,\gamma_I\delta_I}\,c^\dagger_{\alpha_I}\,c^\dagger_{\gamma_I}\,
        c^\dagga_{\delta_I}\,c^\dagga_{\beta_I}\,, 
        \label{eq:PhysHamil}
\eal

where $H^{hop}$ defines the non-local, one-body part of the Hamiltonian, and $H^{loc}_{I}$ are the local terms, including interactions.
We will use capital Latin letters $I,J$ to identify the localized fragments of the system where interactions occur, e.g. atoms in a molecule, and Greek letters $\alpha,\beta,\gamma,\delta$ for generalized orbital indices in the original system.
Within gGut, this Hamiltonian is mapped into a one-body quasiparticle Hamiltonian $H_{qp}$, following

\begin{equation}
    H_{qp} = \sum_{\substack{I\neq J\\ a_Ib_J}}\left(\sum_{\alpha_I\beta_J}\, R^{I,\dagger}_{a_I\alpha_I}\, t_{\alpha_I\beta_J} \,R^{J\dagga}_{\beta_J b_J}-\delta_{IJ}\lambda^{I}_{a_Ib_I}\right)\ d^\dagger_{a_I}\,d^\dagga_{b_I},
    \label{eq:Hqp}
\end{equation}

where we use small Latin letters $a,b$ to differentiate orbitals in the quasiparticle Hilbert space from the orbitals $\alpha,\beta$ in the physical one, and similarly denote quasiparticle creation/annihilation operators as $d$ instead of $c$.
In $H_{qp}$, the interactions are substituted by a renormalization of the one-body hoppings $t$ with the renormalization matrices $R^I$, as well as with local one-body potentials $\lambda^I$.
These $R^I$ and $\lambda^I$ parameters are variationally optimized, within the infinite dimensional limit, following an embedding prescription formulated around the one-body reduced density matrix 
$\langle d^\dagger_{a I} d^\dagga_{b I}\rangle$ (see Ref.~\cite{lanata2015,Mejuto2023a} and the SI). 
The main difference between gGut and the original Gutzwiller Ansatz lies in allowing the quasiparticle Hilbert space to be larger than the physical one. 
Specifically, to each physical orbital $\alpha=1,\dots,2N$ we associate $1+N_g$ auxiliary orbitals, where $N_g> 0$ is denoted as the number of ghosts. The conventional Gutzwiller wavefunction is restored when $N_g=0$.
In other words, in gGut we allow for more $a,b$ auxiliary orbitals than $\alpha,\beta$ physical orbitals.
This can be understood as a way of representing one physical, interacting orbital as a composite of several, non-interacting quasiparticle ones.
After all, one can interpret the new one-body structure in Eq.~\eqref{eq:Hqp} as arising from a substitution $c^\dagger_{\alpha I}\rightarrow \sum_{a} R^{I,\dagger}_{a I, \alpha I}\  d^\dagger_{a I}$.
It is this enlargement of the quasiparticle Hilbert space that allows describing coherent and incoherent features of strong correlation within a one-body picture~\cite{Lanata2017}.

In simple terms, the quasiparticle Hamiltonian in gGut provides us a band structure/molecular orbital picture for correlated electrons.
Directly from $H_{qp}$ we can extract the quasiparticle eigenvalues which enter the topological indices in Eq.~\eqref{N-2}, and furthermore we can find an expression for the Green's function of our original system as

\begin{equation}
    G_{\alpha_I\beta_J}(\omega) = \sum_{ab}R^{I}_{\alpha_Ia_I}
    \Bigg[\frac{1}{\omega+i0^+ - H_{qp}}\Bigg]_{a_Ib_J}\!\!\!\!\!\!R^{J\dagger}_{b_J\beta_J}.
    \label{eq:gGut_GF}
\end{equation}

This expression for the Green's function, carefully justified in Ref.~\cite{Pasqua2025}, can be directly compared with Eq.~\eqref{G vs G*}.
It becomes thus apparent that $H_{qp}$ is playing the role of the quasiparticle Hamiltonian $H_*$ in Eq.~\eqref{H* R}, as its resolvent gives us the quasiparticle Green's function in Eq.~\eqref{G vs G*}.
At this point, it is worth pointing out that the interacting Green's function on the left hand side of Eq.~\eqref{eq:gGut_GF} does not arise from a static, i.e. frequency independent approximation, even if $H_{qp}$ is itself static and the gGut self-consistency involves only the frequency independent density matrix.
The contractions with the renormalization matrices $R^I$ are equivalent to a frequency dependent self-energy, see e.g. Ref.~\cite{Lee2023a}.
The approximation is orbital-local in the $I,J$ fragments, but dynamical in frequency.
Nevertheless, it remains a computationally inexpensive approximation offering a highly interpretable one-body picture for strong correlation, and has consequently found various applications in solid-state~\cite{Lanata2017,Lee2023a,Mejuto2023a,Lee2023b,Giuli2025,Bellomia_intracorr,Tagliente2025,Yao2021,Besserve2022,Chen2025,Sriluckshmy2025,Pasqua2025} and chemical systems~\cite{Mejuto2024}.

Finally, some words are in order regarding the use of gGut on molecular systems.
After all, a typical molecular Hamiltonian does not present the local interaction structure assumed in Eq.~\eqref{eq:PhysHamil}.
In Ref.~\cite{Mejuto2024}, some of us showed how these non-local interactions can be easily recovered by first performing a mean-field decoupling of the non-local interactions, effectively bringing the full molecular Hamiltonian to the form of Eq.~\eqref{eq:PhysHamil}.
The mean-field parameters of this decoupling are just the one-body density matrix elements, and thus they can be self-consistently determined within the gGut formalism.
One arrives thus at a nested self-consistent formalism, with the outer, charge self-consistency fixing the non-local mean-field, and the inner self-consistency resolving the gGut variational problem.
The effect of the non-local interactions on the strong correlated behavior of the system can be thus accounted for.

\subsection{Toy Reactions}

\begin{figure*}
    \centering
    \includegraphics[width=1.\linewidth]{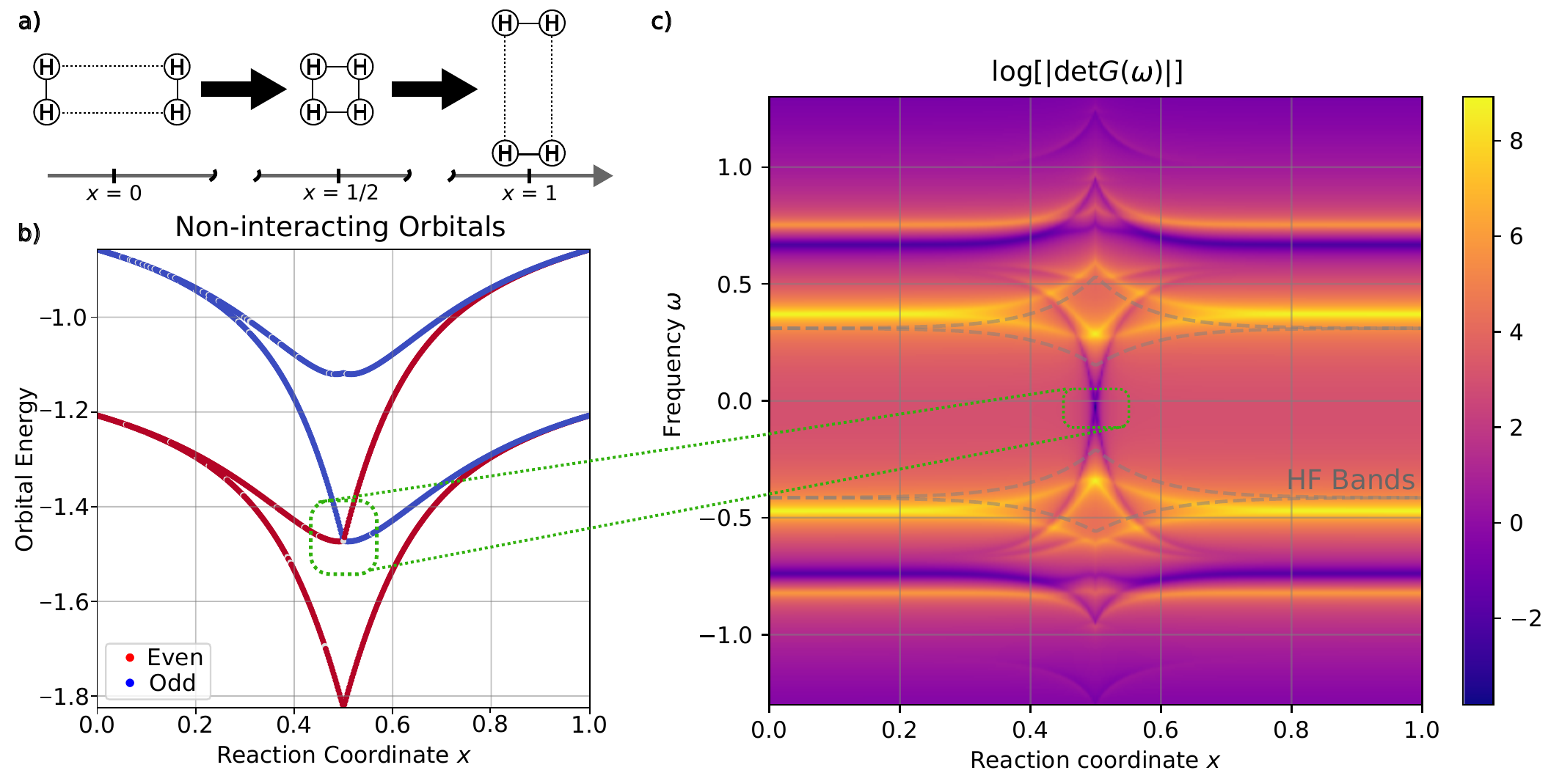}
    \caption{Toy example of Woodward-Hoffmann forbidden reaction for H$_4$ cluster in sto-6g basis. a) Schematic of reaction. b) Non-interacting orbital energies as a function of the reaction coordinate $x$, marking the orbital parity with respect to the $\sigma_{xz}$ mirror plane. c) Logarithm of the absolute value of the determinant of the exact Green's function, with Hartree-Fock orbital energies shown as dashed gray lines. Poles of the green's function appear as bright linese, zeros as dark ones. Note that the crossing of a pair of even/odd non-interacting orbitals at $x = 1/2$ in the corresponds to the crossing of zeros in the interacting Green's function.}
    \label{fig:H4_min}
\end{figure*}

\begin{figure*}
    \centering
    \includegraphics[width=1.\linewidth]{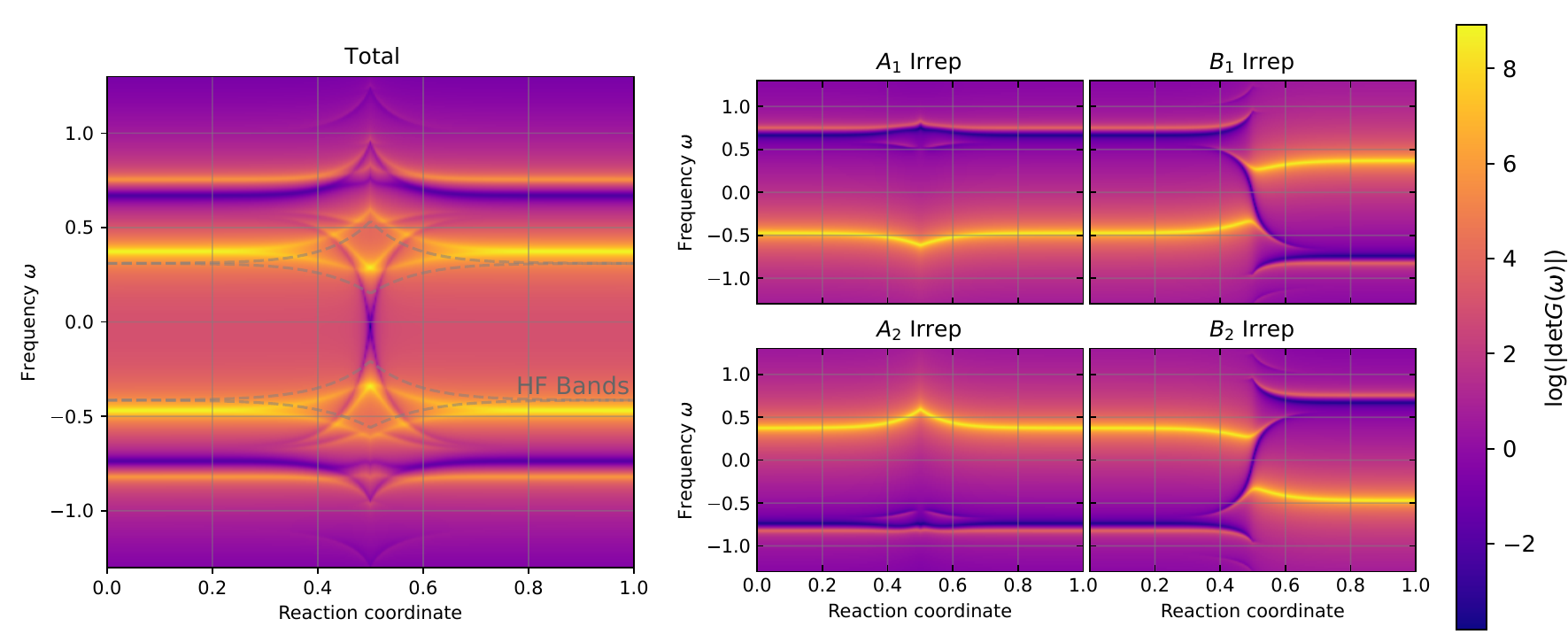}
    \caption{Symmetry analysis of $\mathrm{log}[|\mathrm{det}(G)|]$ for the ED Green's function of the H$_4$ reaction in the sto-6g basis. We compare the total Green's function determinant with its components in the four irreducible representation of the $C_{2v}$ point group.}
    \label{fig:H4_EDSym}
\end{figure*}

To study the phenomenology of the Woodward-Hoffmann rules, we will consider the following toy systems: an H$_4$ rectangle in sto-6g and 6-31g basis and an H$_6$ hexagon in 6-31g basis.
For the case of H$_4$, we will study a prototypical example of Woodward-Hoffmann forbidden reaction, presenting a crossing of non-interacting orbitals of different symmetry at the HOMO-LUMO gap, or equivalently a crossing of zeros of the interacting Green's function.
The simulation in non-minimal 6-31g basis will exemplify how the crossing of orbitals/zeros does not need to take place at $\omega = 0$, as discussed above.
We then proceed to use H$_6$ to show a ``reaction path'' with both Woodward-Hoffmann allowed and forbidden segments.
These two cases will serve to illustrate how a quasiparticle Hamiltonian model achieves a comprehensive description of the relevant topological invariant governing the ``reaction'' in terms of effective, non-interacting orbitals.

We will first illustrate these reactions with the limiting cases of non-interacting orbitals and exact diagonalization (ED), i.e., full configuration interaction (FCI).
Then we will access the quasiparticle description which embodies the best of both limits within ghost Gutzwiller (gGut) embeddings of different numbers of ghosts.
We will be mostly concerned with energetics and the one-body Green's function, or rather its determinant.
The mean-field calculations and computation of Hamiltonian one- and two-body parameters were performed using PYSCF~\cite{PYSCF1,PYSCF2,PYSCF3}. 

\section{The H$_4$ ``Reaction'' - Woodward-Hoffmann Forbidden}

For the H$_4$ system, we propose the following ``reaction'', sketched in panel a of Fig.~\ref{fig:H4_min} and controlled by the reaction coordinate $x$: starting with a horizontal rectangle at $x = 0$, we approach the two (vertical) H$_2$ dimers until a square is formed at $x = 1/2$. Then, we proceed to pull two horizontal H$_2$ dimers apart from the square, until we arrive at a vertical rectangle at $x = 1$.
Before discussing the numerical results, let us begin with a symmetry analysis. 
The H$_4$ planar molecule along the reaction path has $C_{2v}$ symmetry that includes the twofold rotation $C_{2z}$ around the center of mass, and two mirror planes, $\sigma_{xz}$ and $\sigma_{yz}$, orthogonal to the horizontal and vertical axes, respectively. 
The character table is shown in Table~\ref{C2v}. 
\begin{table}[h]
\bealn
\begin{array}{|c|c|c|c|c|}\hline
~ & E &C_{2z} & \sigma_{xz} & \sigma_{yz}\\ \hline
A_1 & 1 & +1 & +1 & +1 \\ \hline
A_2 & 1 & +1 & -1 & -1 \\ \hline
B_1 & 1 & -1 & +1 & -1 \\ \hline
B_2 & 1 & -1 & -1 & +1 \\ \hline
\end{array}
\eal
\caption{Character tables for $C_{2v}$.}
\label{C2v}
\end{table}
At $x=1/2$, the symmetry group raises to $C_{4v}$, which includes also $C_{4z}$. One easily realizes that along the entire path $x\in[0,1]$, including $x=1/2$, the interacting many-body ground state, which contains four electrons due to charge neutrality, is non-degenerate; specifically, a spin-singlet with symmetry $A_1$.\\
We can use a minimal basis of just four molecular orbitals, each of them transforming like one of the four irreps in Table~\ref{C2v}. The advantage of this choice is that the single-particle Green's function 
\eqn{G Matsubara} is automatically diagonal with elements $G_a(i\ep)$, 
$a=A_1,A_2,B_1,B_2$. At $x=1/2$, the two molecular orbitals $B_1$ and $B_2$ transform like the two-dimensional irrep of $C_{4v}$. Therefore, $G_{B_1}(i\ep)$ and $ G_{B_2}(i\ep)$ are equal at $x=1/2$, otherwise they differ. 
Additionally, in the minimal basis, the two molecular orbitals $B_1$ and $B_2$ transform into each other by a particle-hole transformation, thus $G_{B_1}(i\ep)=-G_{B_2}(-i\ep)$. This implies that 
at $x=1/2$
\be\nonumber
G_{B_1}(i\ep)=-G_{B_2}(-i\ep)=-G_{B_1}(-i\ep)\,,
\ee
is purely imaginary. As $\epsilon \to 0$, the Green’s functions either diverge, $G_{B1}(i\epsilon) = G_{B_2}(i\epsilon) \sim 1/i\epsilon$, or vanish, $G_{B1}(i\epsilon) = G_{B_2}(i\epsilon) \sim -i\epsilon$. The former case implies a degenerate ground state, which does not occur. Therefore, the only possibility is that both $G_{B1}(i\epsilon)$ and $G_{B_2}(i\epsilon)$ vanish at $\ep=0$. 
Correspondingly, the eigenvalues 
$\ep_{B_1}(x)$ and $\ep_{B_2}(x)$ of the quasiparticle Hamiltonian 
$H_*(\ep,x)$ at $\ep=0$, which we mentioned may describe either poles or zeros, cross zero energy at $x=1/2$, with one crossing upwards and the other downwards. We must therefore conclude that the ``reaction'' is symmetry-forbidden according to the generalized Woodward-Hoffmann rules proposed by \cite{Xie2025}. \\
Suppose we choose a larger basis than the minimal one. In this case, particle-hole symmetry is lost, and $G_{B_1}(i\ep) \not=-G_{B_2}(-i\ep)$. However, it remains true that $G_{B_1}(0)$ and $G_{B_2}(0)$ will cross zero along the reaction pathway, but now this will occur at different values of $x$. In other words, $\ep_{B_1}(x)$ and $\ep_{B_2}(x)$ do cross zero energy, but at two distinct $x_{B_1}$ and $x_{B_2}$ symmetrically located around $x=1/2$. Since the presence or absence of particle-hole symmetry should not matter as long as the ground state remains unchanged, we must conclude that the transition is still forbidden even when a single Green’s function or, equivalently, a single eigenvalue of the quasiparticle Hamiltonian crosses zero. This aligns with our further generalization of the Woodward-Hoffmann rules. We remark that the existence of two crosses depends on the fact that Luttinger's theorem is obeyed near $x\simeq 0$ and $x\simeq 1$.     

\begin{figure*}
    \centering
    \includegraphics[width=\linewidth]{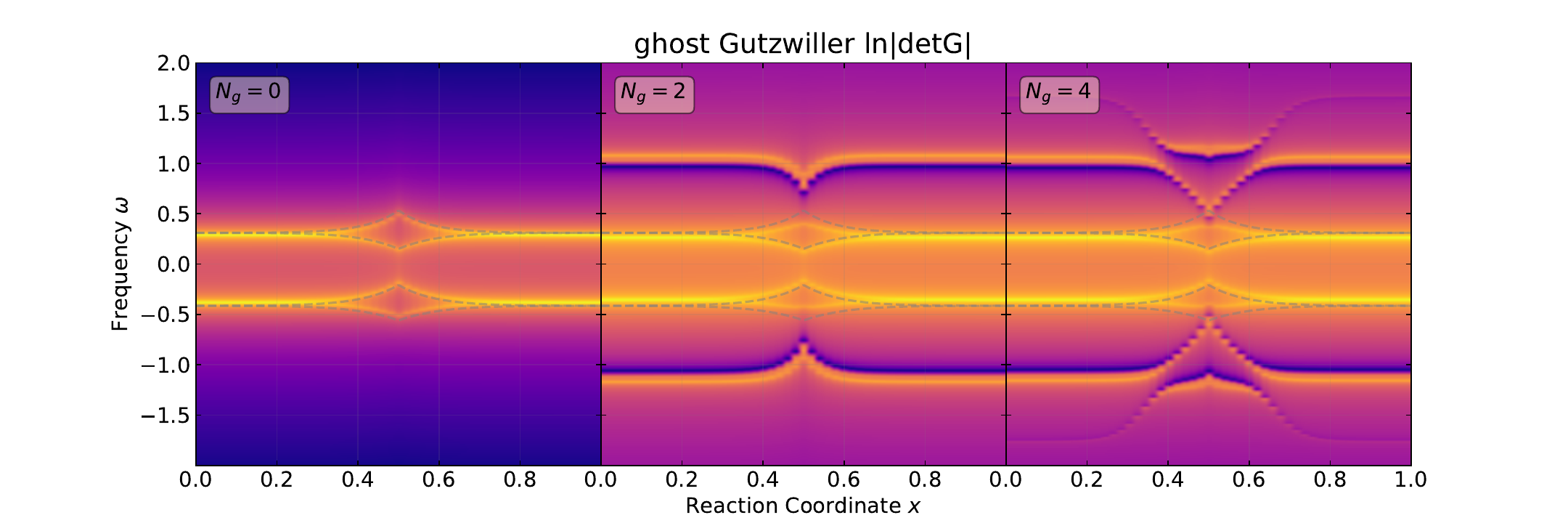}
    \caption{Logarithm of the absolute value of the determinant of the Green's function for the H$_4$ reaction in sto-6g basis, as computed with ghost Gutzwiller with different numbers of ghosts $N_g$. The Hartee-Fock orbital energies are shown as dashed gray lines.}
    \label{fig:H4_min_gGutGF}
\end{figure*}

\subsection{One-body Treatments}

We can first follow the reaction within one-body descriptions, notably the non-interacting molecular orbitals, i.e. the eigenvalues of the one-body part of the H$_4$ Hamiltonian, and a restricted HF mean-field treatment.
The one-body excitations in HF are given by the orbital energies (Koopman's theorem).
These, together with the non-interacting orbital energies, are shown in Fig.~\ref{fig:H4_min}.
Here, the non-interacting orbitals, together with their parity with respect to the $\sigma_{xz}$ mirror plane in the $\mathrm{C_{2v}}$ group, are shown in panel b, while the energies of the Hartree-Fock orbitals are shown as grey dashed lines in panel c.

The orbital energies in panel b of Fig.~\ref{fig:H4_min} are exemplary of the kind of phenomenology we are after:
As proven through symmetry arguments, we have the crossing of odd/even orbitals at $x = \frac{1}{2}$. 
At the HF level this degeneracy is lifted by the interactions, and we have a gapped change of parity in the frontier orbitals (not shown in the figure).

\begin{figure*}
    \centering
    \includegraphics[width=\linewidth]{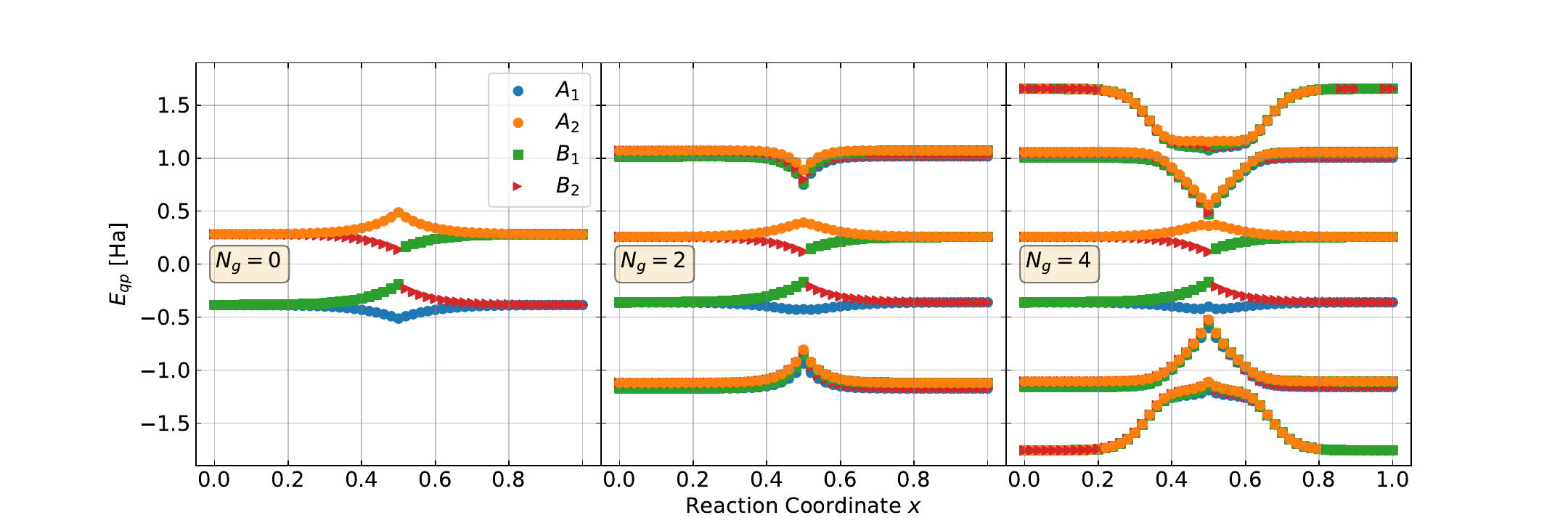}
    \caption{Quasiparticle eigenvalues within the ghost Gutzwiller treatment of the H$_4$ reaction in 6-31g basis, as computed with different numbers of ghosts $N_g$. We indicate the irreducible representation of the $\mathrm{C_{2v}}$ group to which each quasiparticle state corresponds with a different color/marker.}
    \label{fig:H4_min_gGutqp}
\end{figure*}

\begin{figure}
    \centering
    \includegraphics[width=1.0\linewidth]{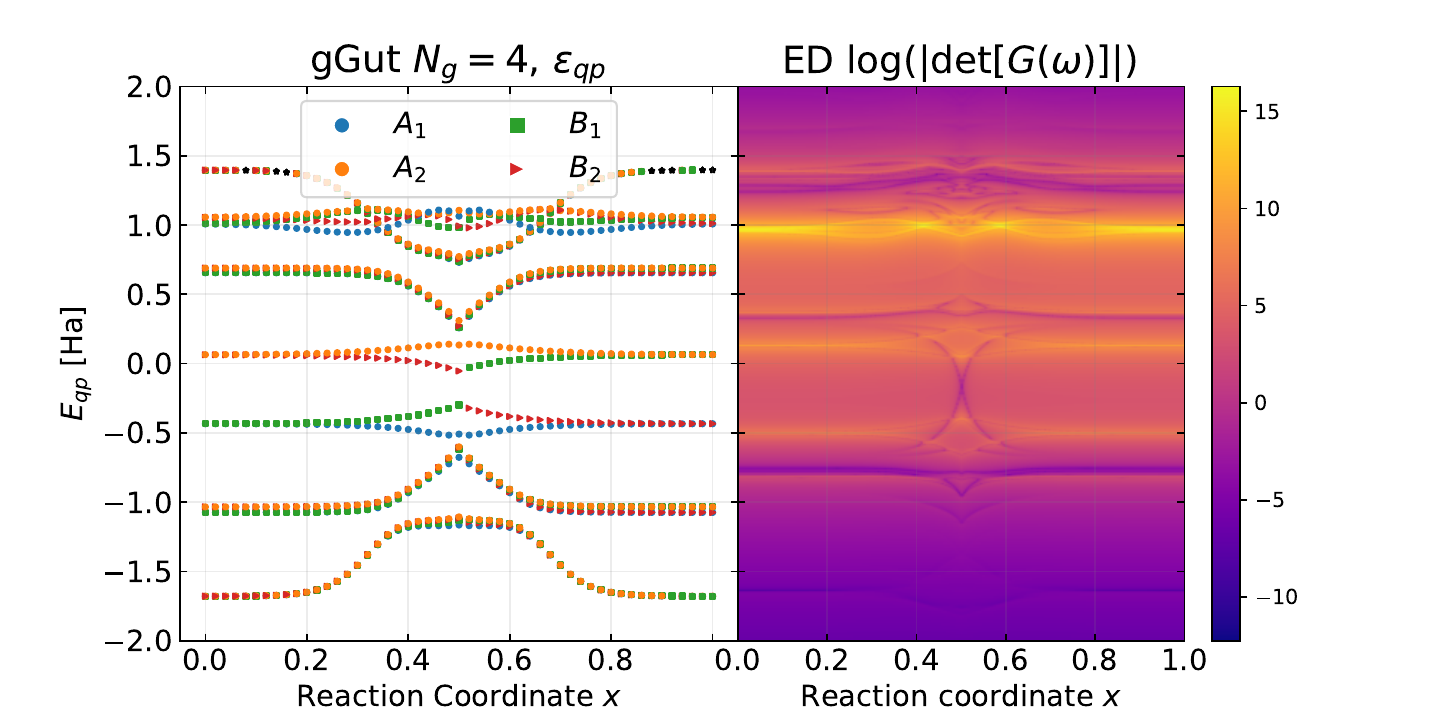}
    \caption{Toy H$_4$ reaction in non-minimal sto-6g basis. Left panel shows the quasiparticle eigenvalues within a ghost Gutzwiller treatment with 4 ghosts per orbital. We indicate the irreducible representation of $\mathrm{C_{2v}}$ to which each quasiparticle state corresponds with a different color/marker. Right panel show the logarithm of the absolute value of the exact determinant. Note that in this non-minimal basis model, the crossing of zeros does not happen at zero energy.}
    \label{fig:H4_non_min}
\end{figure}

\subsection{Exact Treatment}

We can compare the non-interacting orbital energies in panel b of Fig.~\ref{fig:H4_min} with the poles and zeros of the determinant of the exact one-body Green's function in panel c.
We note that the main poles around the Fermi level closely follow the HF orbital energies, yet there is an important addition: the presence of two Green's function zeros crossing the Fermi level at $x=\frac{1}{2}$, just like the non-interacting frontier orbitals do in the b panel of Fig.~\ref{fig:H4_min}.
We note in passing that the ED Green's function has several additional zeros at higher energy, and that it seems that the presence of the zeros close to the Fermi level at $x = \frac{1}{2}$ is repelling the poles, essentially deforming the curvature of the frontier orbital poles.

Our toy system recovers the picture in which the Green's function zeros seem to take on the role of the non-interacting orbital energies for the purposes of the Woodward-Hoffmann rules, as first suggested in Ref.~\cite{Xie2025}.
The role of the zeros in the Woodward-Hoffmann setting can be made even more explicit by looking at the symmetry blocks of the exact Green's function separately, as shown in Fig.~\ref{fig:H4_EDSym}.
Indeed, since the Green's function can be diagonalized in the basis of irreducible representations, the logarithm of its determinant is just a sum of the contributions of each symmetry block.
In Fig.~\ref{fig:H4_EDSym} we can see how, indeed, the zeros crossing at $x = \frac{1}{2}$ correspond to the $B_1$ and $B_2$ irreducible representations of $C_{2v}$.
Moreover, we see that this crossing of zeros coincides with a degeneracy between two pairs of poles in the $B_1$ and $B_2$ sectors on both sides of $\omega = 0$.
Thus, a suggestive connection between the ED and non-interacting formulations of the Woodward-Hoffmann rules arises: the non-interacting orbitals, upon the introduction of interactions, experience in general a shift of their  corresponding Green's function poles.
In the case of strong correlation, as in our H$_4$ model around $x = \frac{1}{2}$, these poles end up splitting into Hubbard-like high-energy ones, with the concurring appearance of in-gap Green's function zeros. The crossing between HOMO, $B_1$ symmetry, and LUMO, $B_2$ symmetry, which occurs at $x=1/2$ within the independent-particle approximation, turns, in the fully-interacting approach, into a HOMO-LUMO symmetry transmutation without any gap closing. Precisely at $x=1/2$, both HOMO and LUMO are twofold degenerate, while at any $x\not=1/2$ they are non-degenerate. Correspondingly, the Green's function zeros of $B_1$ and $B_2$ symmetry cross at $x=1/2$.
This offers a reasonable justification of why the zeros can take the role of the non-interacting orbitals when evaluating the Woodward-Hoffmann rules.

Now, while instructive, an ED treatment is in general non-scalable beyond minimal basis representations of small molecules.
Furthermore, while the justification in terms of Hubbard bands was successful in H$_4$, this will not necessarily be applicable to a broad range of molecular systems.
It would be desirable to find a flexible description, akin to a molecular orbital theory, that still is rigorously rooted in the interacting electron limit.
In the following section, we will show how a quasiparticle treatment of the interactions recovers precisely such a picture, formulating both zeros and poles of an interacting Green's function in terms of effectively non-interacting orbital energies.

\begin{figure*}
    \centering
    \includegraphics[width=1.\linewidth]{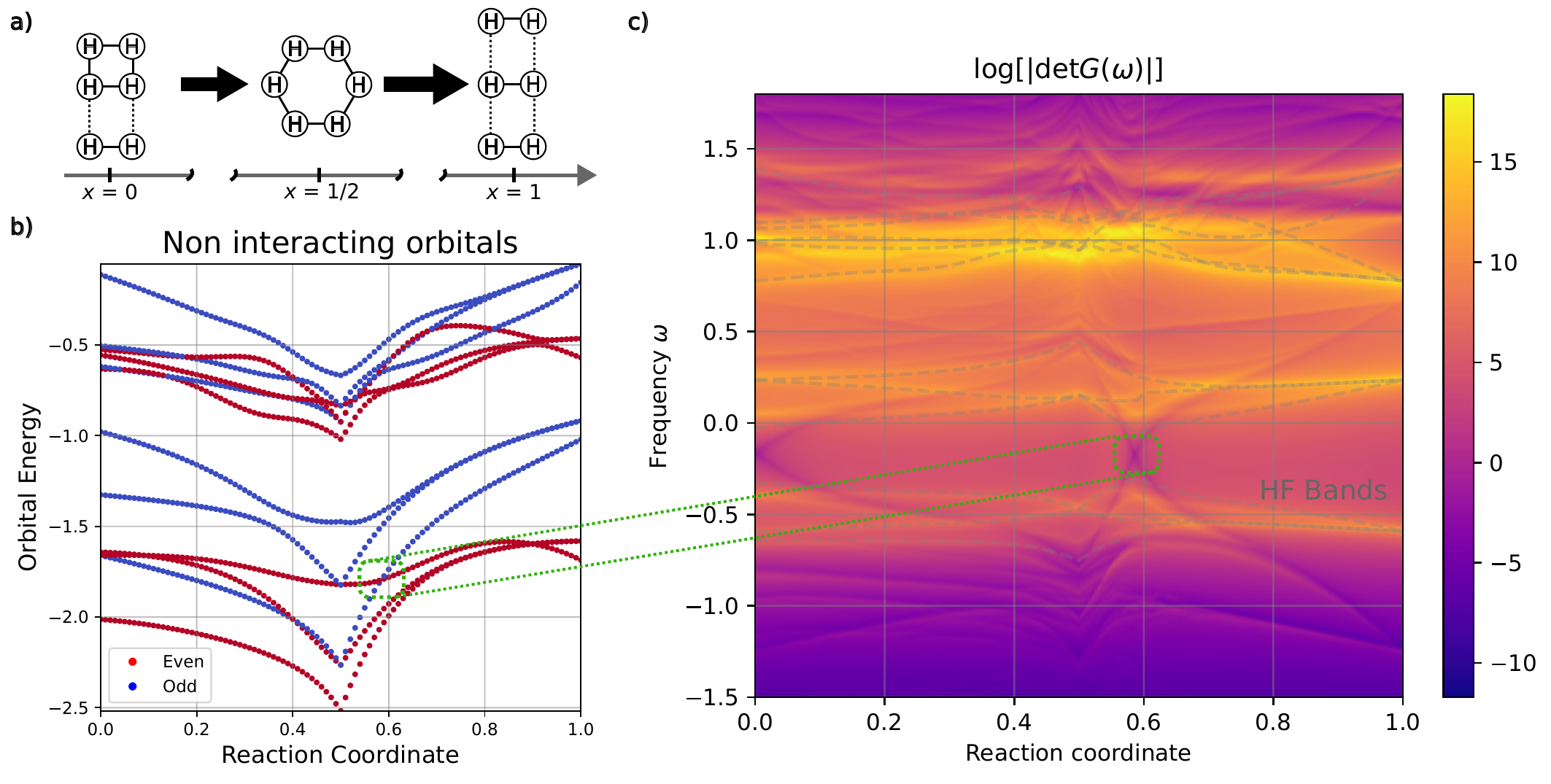}
    \caption{Toy example presenting both Woodward-Hoffmann allowed and forbidden reactions for H$_6$ cluster in 6-31g basis. a) Schematic of reaction. The first half $x = 0\rightarrow 1/2$ is symmetry-allowed, while the second half $x= 1/2 \rightarrow 1$ is symmetry forbidden. b) Non-interacting orbital energies as a function of the reaction coordinate $x$, marking the orbital parity with respect to the $\sigma_{zx}$ rotation. c) Logarithm of the absolute value of the determinant of the exact Green's function, with Hartree-Fock orbital energies shown as dashed gray lines. Poles of the green's function appear as bright linese, zeros as dark ones. Note that the crossing of a pair of even/odd non-interacting orbitals in the range $x \in (1/2,1)$ in the corresponds to the crossing of zeros in the interacting Green's function.}
    \label{fig:H6_min}
\end{figure*}

\begin{figure*}
    \centering
    \includegraphics[width=1.\linewidth]{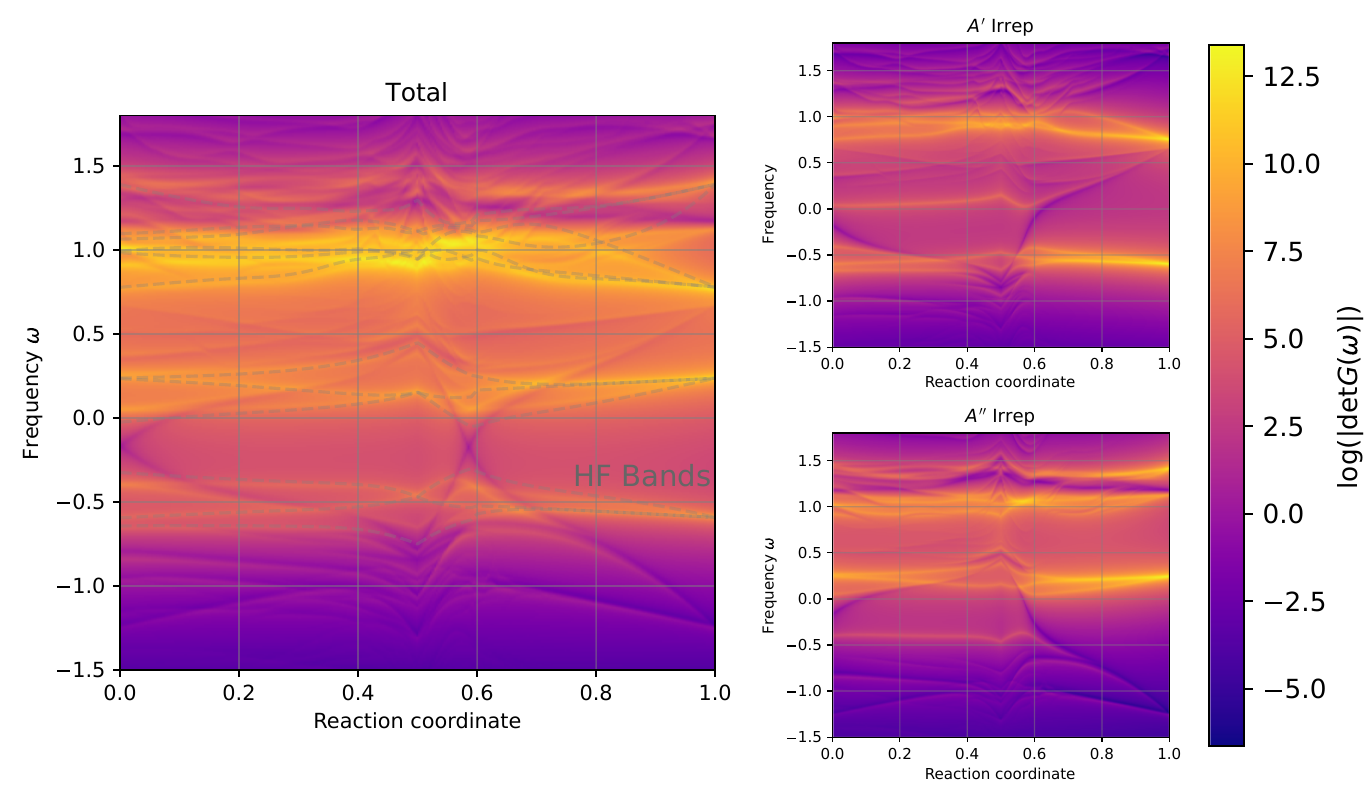}
    \caption{Symmetry analysis of $\mathrm{log}[|\mathrm{det}(G)|]$ for the ED Green's function of the H$_6$ reaction in the 6-31g basis. We compare the total Green's function determinant with its components in the two irreducible representation of the $C_{s}$ point group.}
    \label{fig:H6_EDSym}
\end{figure*}

\subsection{Quasiparticle Embedding}

We perform gGut simulations on this system.
We will consider an atomic embedding, i.e. treating each H atom with its own impurity model.
Embedding more than one H atom together is possible, but does not change the conclusions of our study.
In Fig.~\ref{fig:H4_min_gGutGF}, we show the determinant of the Green's function for simulations with no ghosts ($N_g = 0$, leftmost panel), 2 ghosts per impurity ($N_g = 2$, center panel) and 4 ghosts per impurity ($N_g = 4$, rightmost panel).

The progression from $N_g = 0$ to $N_g = 4$ in Fig.~\ref{fig:H4_min_gGutGF} shows nicely how gGut interpolates between a mean-field and correlated treatment by increasing the number of ghosts.
Indeed, while the Green's function with no ghosts on the leftmost panel essentially reproduces the mean-field Hartree-Fock orbital energies (shown as gray dashed lines), adding ghosts modifies slightly these energies and introduces the presence of Green's function zeros.
Going from $N_g = 2$ to $N_g = 4$ adds features mostly at higher energies, a common behavior in embedding-like approximations~\cite{Mejuto2020,Florez2025}, and we note that the region that exhibits the most changes is around $x = 1/2$, precisely the most correlated region of the ``reaction''.
Now, while the addition of ghosts qualitatively reproduces most of the main features of the exact Green's function in panel c of Fig.~\ref{fig:H4_min}, e.g. the satellite poles and zeros at higher energy, our gGut results never quite show the zeros crossing at $x = 1/2$.
They get closer from $N_g = 2$ to $N_g = 4$, but never quite touch.
This, however, has nothing to do with the approximations in gGut, but rather with the mean-field Hartree-Fock that precedes it.
Indeed, we perform a restricted Hartree-Fock calculation to generate the effective local Hamiltonian model to be embedded with gGut, following the prescription of Ref.~\cite{Mejuto2024}, but this introduces a symmetry breaking at the square geometry.
A fully symmetric solution would require the degenerate $B_1$ and $B_2$ orbitals of the square H$_4$ to be each singly occupied, yet this is not possible in restricted Hartree-Fock.
This asymmetry remains in the solution of the one-body quasiparticle Hamiltonian in Eq.~\eqref{eq:Hqp}, and hence persists in the gGut treatment.
As a consequence, the H$_4$ model we embed has lower symmetry than the true molecule, and the zeros remain gapped, similarly how the restricted Hartree-Fock orbital energies are gapped themselves.
This issue persists even after converging the external, charge self-consistency.
We can resolve it by using a multireference, two-determinant solution to build the ``mean-field'' density matrix defining the embedding problem, which indeed leads to the zeros crossing at H$_4$.
Nevertheless, as our analysis of the quasiparticle eigenvalues will show in the next paragraph, this is not necessary to capture the essence of the Woodward-Hoffmann rules.

Indeed, instead of focusing on the position of the zeros in the fully interacting Green's function, we can turn our attention to the negative eigenvalues of the quasiparticle Hamiltonian.
Following Eq.~\eqref{N-2}, the topological index marking whether a reaction complies with the Woodward-Hoffmann rules~\cite{Xie2025} can be simply evaluated by counting the number of negative eigenvalues in each irreducible representation.
If this number changes between the reactants and the products along a reaction, this is Woodward-Hoffmann forbidden, otherwise it is allowed.
Hence, we turn our attention to the quasiparticle spectrum within the ghost Gutzwiller approximation, shown in Fig.~\ref{fig:H4_min_gGutqp} for the gGut simulations of different number of ghosts $N_g$.
Here, the marker/color combination for each quasiparticle energy indicates the irreducible representation of $C_{2v}$ to which it belongs.
Paying attention to the orbitals around energy zero, we see that for $x < 1/2$ the HOMO orbital transforms as $B_1$ (green), whereas for $x > 1/2$ it transforms as $B_2$ (red).
In other words, at $x = 1/2$ there is a change of the topological index in Eq.~\eqref{N-2} for the $B_{1}$ and $B_2$ representations, identifying the reaction as Woodward-Hoffmann forbidden.
If we carefully ensured the preservation of the full $C_4$ symmetry at the $x = 1/2$ point, we would observe the $B_1$ and $B_2$ orbital energies exactly cross at this point, perfectly recovering the simple picture present in the non-interacting orbitals (panel b of Fig.~\ref{fig:H4_min}), but here from a rigorous, interacting, many-body treatment.
As was the case with the fully interacting Green's functions in Fig.~\ref{fig:H4_min_gGutGF}, we note that adding ghosts does not fundamentally change the behaviour of the quasiparticle orbitals around the HOMO-LUMO gap, but instead just adds additional states at higher energies.

Finally, it is worth discussing that this crossing of orbitals of different symmetry at $x = 1/2$ does not need to occur exactly at zero energy.
Indeed , already in this minimal basis calculation, where particle-hole symmetry is not perfectly preserved, this is not quite the case.
The exact results in panel c of Fig.~\ref{fig:H4_min} show that the crossing of Green's function zeros, completely equivalent to the crossing of quasiparticle orbitals, happens at slightly negative energies.
This becomes even more pronounced in a non-minimal basis calculation.
To exemplify this, we present in Fig.~\ref{fig:H4_non_min} the exact Green's function determinant and the quasiparticle energies from a $N_g = 4$ gGut calculation for the same H$_4$ reaction as in panel a of Fig.~\ref{fig:H4_min}, but with a 6-31g basis, i.e., a basis with 1 1s orbital and 1 2s orbital for each H atom.
In this basis, we find exactly the same phenomenology as with the minimal basis results, with the only noticeable differences being the notable shift of the crossing of Green's function zeros to more negative energies, as well as the presence of the 2s excitation manifold at higher energies.
Note further that, around $x = 1/2$, a few additional quasiparticle states become negative.
As discussed in the theory section, this lack of conservation of the number of negative energy quasiparticle states may be related to deviations from Luttinger's theorem, well established in the solid state literature~\cite{Jan} but less explored for molecules.
Pursuing the deeper meaning, if any, of such a deviation presents an exciting direction for future work.

\section{The H$_6$ ``Reaction'' - A Model with Forbidden and Allowed Directions}

For our second example, we choose a ``reaction'' in an H$_6$ system in 6-31g basis, schematically represented in panel a of Fig.~\ref{fig:H6_min}.
We start with an H$_4$ square and an H$_2$ dimer, initially distant and approaching each other along the vertical direction, forming a perfect H$_6$ hexagon, and finally dissociating into three vertically stapled H$_2$ dimers.
The reaction is monitored by a reaction coordinate $x$, such that again at $x = 1/2$ we have the most symmetric configuration, here the hexagon.
Details of the limiting geometries ($x = 0, 1/2, 1)$) can be found in the SI, the reaction develops as a linear interpolation between $x = 0,1/2$ first, and then between $x = 1/2,1$.

As with the H$_4$ case, one can perform a symmetry analysis of the reaction. 
For any $x\not=1/2$, the planar molecule has $C_{s}$ symmetry, which enlarges to $C_{6v}$ at $x=1/2$. 
Unlike in the H$_4$ case, this does not enforce a crossing of orbitals of different symmetry at $x = 1/2$, so the ``reaction'' could, potentially, be Woodward-Hoffmann allowed.
However, upon inspection of the nodal structure of the molecular orbitals of the limiting ($x = 0, 1/2, 1$) configurations, it becomes apparent that the situation can be a bit richer than this:
the nodal structure of the H$_6$ hexagon requires using, at least, one anti-bonding orbital for one of the three H$_2$ dimers at $x = 1$.
Given that bonding and anti-bonding orbitals in H$_2$ have opposite parity, the second half should be Woodward-Hoffmann forbidden, while \emph{a priori} the first part can be allowed!
It is worth stressing at this point that, unlike in the H$_4$ case, the molecular geometry along the reaction does not impose any symmetry-driven constraint on the position of the orbital crossing in the $x\in(1/2,1]$ region.
Instead, the point at which they cross will correspond to some accidental degeneracy, arising from the nature of reactant and product.
We now proceed to thoroughly analyze this situation numerically.

\subsection{One-body Treatments}

Once again, we start by analyzing the ``reaction'' from a one-body perspective.
In panel b of Fig.~\ref{fig:H6_min}, we show the non-interacting orbital energies of the H$_6$ system, marking the parity of each orbital with respect to the $\sigma_{zx}$ plane of the $C_{s}$ group with color.
For the first half of the reaction, we observe that there is no crossing between orbitals of opposite parity in the HOMO-LUMO gap, so indeed the $x\in[0,1/2)$ reaction should be Woodward-Hoffmann allowed.
On the other hand, as anticipated through symmetry arguments, the second half of the reaction presents a crossing between an even and odd orbitals at the HOMO-LUMO gap close to $x\sim 0.6$, signaling it as Woodward-Hoffmann forbidden.
As it stands, this toy ``reaction'' is a great example to test our theory, as it has both an allowed and a forbidden component.

Just like in the H$_4$, the mean-field treatment with restricted Hartree-Fock leads to gapped orbital energies across the full ``reaction'' pathway (see dashed lines in panel c of Fig.~\ref{fig:H6_min}).
This level repulsion follows from interactions, and we hence shift our attention to exact descriptions of the system next.

\begin{figure*}
    \centering
    \includegraphics[width=1.\linewidth]{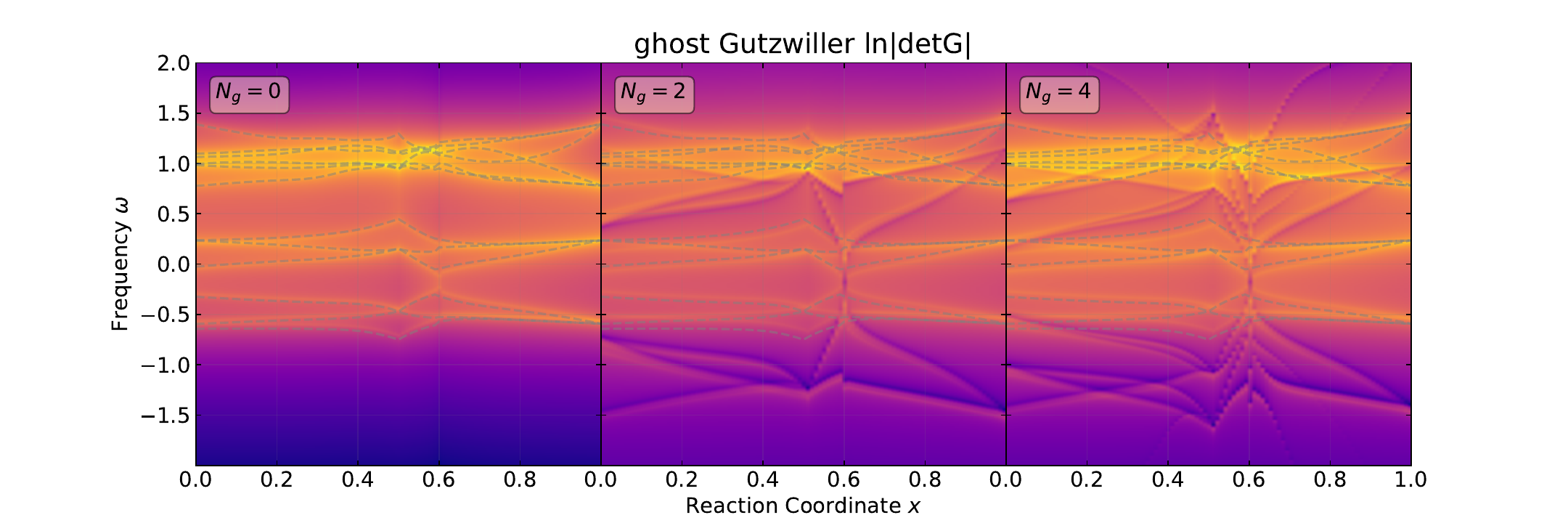}
    \caption{Logarithm of the absolute value of the determinant of the Green's function for the H$_6$ reaction in 6-31g basis, as computed with ghost Gutzwiller with different numbers of ghosts $N_g$. The Hartee-Fock orbital energies are shown as dashed gray lines.}
    \label{fig:H6_min_gGutGF}
\end{figure*}

\begin{figure*}
    \centering
    \includegraphics[width=\linewidth]{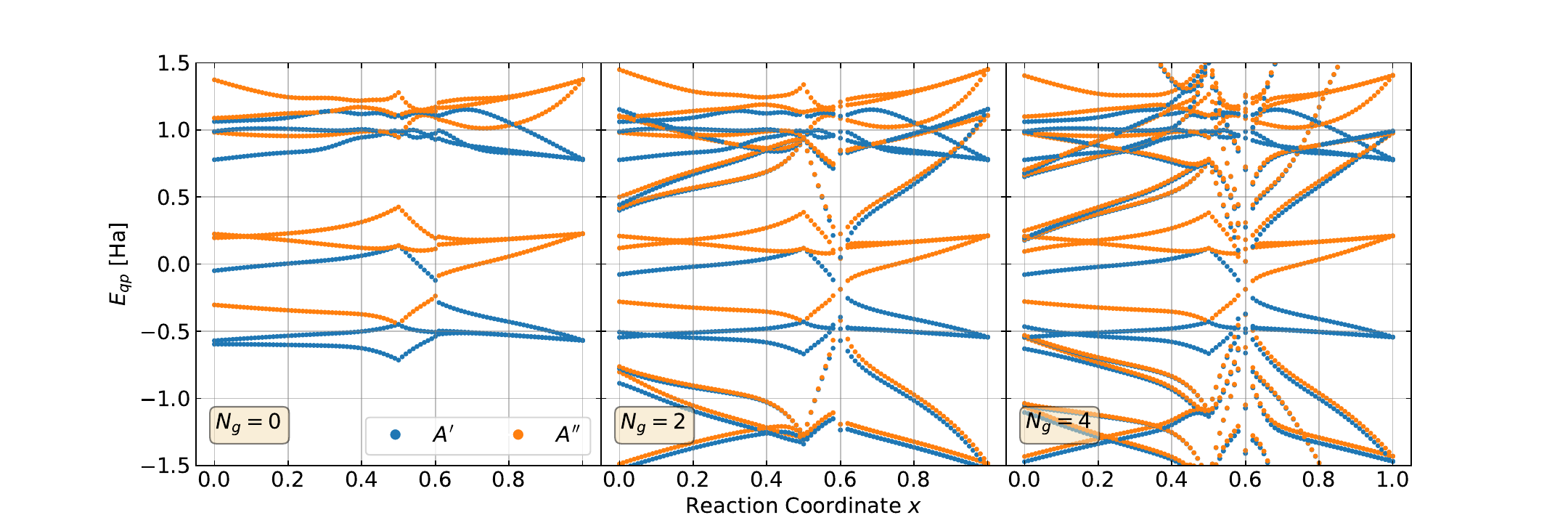}
    \caption{Quasiparticle eigenvalues within the ghost Gutzwiller treatment of the H$_6$ reaction in 6-31g basis, as computed with different numbers of ghosts $N_g$. We indicate the irreducible representation of the $\mathrm{C_{s}}$ group to which each quasiparticle state corresponds with a different color/marker.}
    \label{fig:H6_min_gGutqp}
\end{figure*}

\subsection{Exact Treatment}

We turn now to the exact solution of our H$_6$ system.
In panel c of Fig.~\ref{fig:H6_min} we show the logarithm of the absolute value of the determinant of the interacting Green's function, which we can compare directly with the Hartree-Fock orbital energies (dashed lines) and the non-interacting orbital energies (panel b in Fig.~\ref{fig:H6_min}).
The poles of the Green's function (bright lines), which represent the single particle excitations, follow qualitatively the Hartree-Fock energies, modulo a noticeable increase of the HOMO-LUMO gap.
The most notable difference in terms of poles is the presence of faint, satellite peaks at higher energies.

Notwithstanding this small difference, as with H$_4$ the poles do not give us any obvious information on the compliance of the Woodward-Hoffmann rules.
For this, it is more useful to observe the behavior of the Green's function zeros (dark lines in panel c of Fig.~\ref{fig:H6_min}).
Here, there is a crossing of zeros inside the HOMO-LUMO gap essentially at the same place where the non-interacting frontier orbitals of different parity cross (see panel b of Fig.~\ref{fig:H6_min}).
Meanwhile, no crossing of zeros is apparent for the first half of the ``reaction'' ($x < 1/2$), other than at the starting point $x = 0$, due to the enhanced symmetry of the nearly isolated H$_4$ plaquette.
We recover thus the phenomenology proposed in Ref.~\cite{Xie2025}, and observe how the Green's function zeros adopt the role of the non-interacting orbitals with respect to the Woodward-Hoffmann rules.

Just as in the case of the H$_4$ ``reaction'', we can analyze the poles and zeros in terms of the irreducible representations of the reaction's symmetry, in this case $C_{s}$.
In Fig.~\ref{fig:H6_EDSym}, we show how each of the crossing zeros corresponds to one of the two irreducible representations.
At $x\approx 0.6$, these zeros become degenerate, at the same time as two sets of poles, one above and the other below the crossing zeros.
We recover thus the same picture as in H$_4$: the crossing of zeros seems generated by the degeneracy of two sets of Hubbard-like poles, which one can consider emerging from the corresponding non-interacting orbitals in Fig.~\ref{fig:H6_min}b upon the presence of interactions.

\subsection{Quasiparticle Embedding}

We finally turn to the gGut treatment of the H$_6$ ``reaction''.
We perform an atomic embedding, i.e., each atom has it's own impurity model, with three different numbers of ghosts $N_g = 0, 2, 4$ per embedded atom.
We note that, unlike in the H$_4$ calculations, it is actually possible to obtain a proper crossing of zeros in the gGut solution of the H$_6$ reaction by performing the full charge self-consistency.
This stems from the fact that the zero crossing originates from an accidental degeneracy around $x = 0.6$, rather than an actual symmetry of the Hamiltonian.
Thus, even if the original Hartree-Fock guess presents no degenerate orbitals, the charge self-consistency can drive the quasiparticle orbital energies to become degenerate.

In Fig.~\ref{fig:H6_min_gGutGF} we show the logarithm of the absolute value of the determinant of the gGut Green's function for the simulations with different numbers of ghosts.
It is again apparent how tweaking the number of ghosts allows us to seamlessly interpolate between the Hartree-Fock orbital energies (dashed lines in panel c of Fig.~\ref{fig:H6_min}) and the exact Green's function (heat map in panel c of Fig.~\ref{fig:H6_min}), including satellite poles and the presence of zeros.
Importantly, once we include ghosts (middle and right panel in Fig.~\ref{fig:H6_min_gGutGF}), the charge self-consistency introduces a proper crossing of zeros at $x\approx 0.6$.
This is accompanied with a slight discontinuity in the energy of the high-energy poles and zeros.

These features can again be analyzed in term of the effectively non-interacting quasiparticle orbital energies.
We show these, color/marker coded to indicate their corresponding $C_{s}$ irreducible representation, in Fig.~\ref{fig:H6_min_gGutqp}.
While there are more orbitals to keep track of than in the H$_4$ case, particularly for larger numbers of ghosts, the phenomenology is actually quite transparent.
By focusing on the symmetry character of the orbitals around energy zero, we observe how the frontier orbitals, pertaining to different irreducible representations $A'$ and $A''$, cross at $x \approx 0.6$.
This is of course nothing else than the change of the topological index in Eq.~\eqref{N-2} which encodes the Woodward-Hoffmann rules.
Seeing how such a change in the index of any of the irreducible representations does not take place in the first half of the ``reaction'', we confirm from the quasiparticle picture that the $x\in[0,1/2)$ component is Woodward-Hoffmann allowed, while the $x\in(1/2,1]$ component is forbidden.
Our quasiparticle picture thus elegantly recovers and justifies the non-interacting intuition behind these well established rules, endowing them with a rigorous basis that holds in the fully interacting limit. 

\section{Conclusion}

In this work we have provided a rigorous justification for how the Woodward-Hoffmann rules, originally formulated in terms of non-interacting molecular orbitals, can actually work for  interacting systems.
To this end, we have composed a comprehensive formalism bridging between the original non-interacting picture and a recently proposed fully interacting description in terms of Green's function zeros~\cite{Xie2025}, using the notion of the quasiparticle Hamiltonian.
This object can be used to propose simplified descriptions of many correlated phenomena in terms of effective, non-interacting particles, while having a formal origin in the fully interacting one-particle Green's function.
We have provided a derivation of this quasiparticle Hamiltonian which does not invoke at any point the self-energy, avoiding thus any assumptions on its functional form, which could limit the applicability of the derivation.
Moreover, within the ghost Gutzwiller (gGut) formalism we have provided an inexpensive and accurate strategy to access this quasiparticle Hamiltonian in molecular systems.
All together, this results in a reformulation of the theoretical underpinnings of the Woodward-Hoffmann rules which does not need any non-interacting approximation, together with a computational recipe to implement this formulation to the analysis of chemical reactions. 
While, arguably, a Hartree-Fock description is often sufficient for the organic molecules to which the Woodward-Hoffmann rules are typically applied, the gGut framework used here offers the advantage of working both in the weak and strongly correlated regimes.
Hence, this paints a promising picture for applying the quasiparticle analysis in this paper to other families of molecules.

We exemplify our formulation along two toy ``reactions'' which, while not making any claims or aspirations of explicit chemical realism, concisely collect all main ingredients to showcase the Woodward-Hoffmann phenomenology.
On the one hand, a stereotypical example of a Woodward-Hoffmann forbidden process in a H$_4$ ``reaction'', and on the other a H$_6$ composed of both an allowed and a forbidden part.
Comparing the non-interacting, exact, and quasiparticle descriptions of the ``reaction'', it becomes evident how the latter describes all relevant features present in the fully interacting case within an effective, non-interacting representation.
Furthermore, comparing the H$_4$ scenario between a minimal and non-minimal basis, we unveil apparent violations of Luttinger's theorem in a molecular system, a facet of electron correlation so far unexplored in these systems.
It could be interesting to further investigate the role of the Luttinger integral in chemical settings.

Our work shows how an effective quasiparticle perspective of electron correlation can connect the complex, interacting chemical reality with intuitive, phenomenological rules-of-thumb. 
Such a perspective can be rigorously derived from first principles and systematically accessed with embedding-type approximations. 
The quasiparticle picture holds thus great potential for establishing new guidelines and rules-of-thumb for complex chemical systems of technological interest, particularly those where strong electron correlation plays a leading role, such as transition metal-based complexes.

\section*{Acknowledgments}
We gratefully acknowledge engaging and insightful discussions with Fábris Kossoski, Pierre-François Loos, Lukas M\"uchler and Ivan Pasqua.
This study has been partially supported through the EUR grant NanoX n° ANR-17-EURE-0009 in the framework of the ``Programme des Investissements d'Avenir''.


\appendix
\section{ghost Gutzwiller Equations}

Here we summarize the main equations of the ghost Gutzwiller approximation, particularly when applied to an \emph{ab initio} molecular model.
Details can be found in Ref.~\cite{Mejuto2024}.
We start with a molecular Hamiltonian of the form

\beal
    H_{mol} = \sum_{\alpha\beta\ \sigma} t_{\alpha \beta}\  c^\dagger_{\alpha\sigma} c^\dagga_{\beta\sigma} + \frac{1}{2}\sum_{\substack{\alpha\beta\gamma\delta\\ \sigma\sigma'}} U_{\alpha\beta\,\gamma\delta}\  c^\dagger_{\alpha\sigma} c^\dagger_{\gamma\sigma'}c^\dagga_{\delta\sigma'}c^\dagga_{\beta\sigma},
    \label{eq:SI_Hmol}
\eal

where $c^\dagger$ denote physical creation/annihilation operators, Greek letters indicate physical orbitals, we have explicitly separated the spin degree of freedom $\sigma,\sigma'$, and $t$ and $U$ are the one- and two-body components of the energy respectively.
To apply the ghost Gutzwiller approximation we have to first divide the molecular orbitals into correlated fragments.
Each such fragment represents a subset of orbitals which are strongly correlated between each other, but weakly correlated with orbitals from other fragments.
We will label these fragments with capital Latin letters $I,J$, and reserve $Y$ for those orbitals that are not expected to be correlated, and which thus need not be included in any correlated fragment.
Upon this organization of orbitals in fragments, we can rewrite the Hamiltonian in Eq.~\eqref{eq:SI_Hmol} as

\beal
    H_{mol} &= \sum_I H^{loc}_{I} + H^{loc}_Y + H^{hyb},\\
    H^{loc}_{I} &= \sum_{\alpha_I\beta_I,\sigma}t_{\alpha_I\beta_I}\ c^\dagger_{\alpha_I\sigma}c^\dagga_{\beta_I\sigma}\\
    &+\frac{1}{2}\sum_{\substack{\alpha_I\beta_I\gamma_I\delta_I\\ \sigma\sigma'}}U_{\alpha_I\beta_I\,\gamma_I\delta_I}\ c^\dagger_{\alpha_I\sigma}c^\dagger_{\gamma_I\sigma'}c^\dagga_{\delta_I\sigma'}c^\dagga_{\beta_I\sigma},\\
    H^{loc}_Y &= \sum_{\alpha_Y\beta_Y,\sigma}t_{\alpha_Y\beta_Y}\ c^\dagger_{\alpha_Y\sigma}c^\dagga_{\beta_Y\sigma}\\
    &+\frac{1}{2}\sum_{\substack{\alpha_Y\beta_Y\gamma_Y\delta_Y\\ \sigma\sigma'}}U_{\alpha_Y\beta_Y\,\gamma_Y\delta_Y}\ c^\dagger_{\alpha_Y\sigma}c^\dagger_{\gamma_Y\sigma'}c^\dagga_{\delta_Y\sigma'}c^\dagga_{\beta_Y\sigma},\\
    H^{hyb} &= \sum_{I\neq J}H^{hyb}_{IJ} + \sum_I H^{hyb}_{IY},
    \label{eq:SI_Mol_frag}
\eal

where $H^{loc}_I$ collects all terms local to fragment $I$, $H^{loc}_Y$ collects all terms involving exclusively uncorrelated orbitals, and $H^{hyb}$ collects all other terms, split into $H^{hyb}_{IJ}$ denoting all terms involving at least two correlated fragments $I,J$ and $H^{hyb}_{IY}$ denoting all terms involving just one correlated fragment $I$ and at least one uncorrelated orbital from $Y$.
Once this fragmentation of the Hamiltonian has been carried out, we can bring Eq.~\eqref{eq:SI_Mol_frag} to the form of Eq.~\eqref{eq:PhysHamil} by performing a mean-field decoupling of all interaction terms not present in any $H^{loc}_I$.
Essentially, we carry out a mean-field decoupling following

\begin{widetext}
\beal
    c^{\dagger}_{\alpha\sigma}c^\dagger_{\gamma\sigma'}c^\dagga_{\delta\sigma'}c^\dagga_{\beta\sigma} \rightarrow\  &c^\dagger_{\alpha\sigma}c^\dagga_{\beta\sigma}\langle c^\dagger_{\gamma\sigma'}c_{\delta\sigma'}\rangle+c^\dagger_{\gamma\sigma'}c^\dagga_{\delta\sigma'}\langle c^\dagger_{\alpha\sigma}c_{\beta\sigma}\rangle\\
    &-\delta_{\sigma\sigma'}\left[c^\dagger_{\alpha\sigma}c^\dagga_{\delta\sigma}\langle c^\dagger_{\gamma\sigma}c_{\beta\sigma}\rangle+c^\dagger_{\gamma\sigma}c^\dagga_{\beta\sigma}\langle c^\dagger_{\alpha\sigma}c_{\delta\sigma}\rangle\right]\\
    &-\langle c^\dagger_{\alpha\sigma}c^\dagga_{\beta\sigma}\rangle\langle c^\dagger_{\gamma\sigma'}c_{\delta\sigma'}\rangle+\delta_{\sigma\sigma'}\langle c^\dagger_{\alpha\sigma}c^\dagga_{\delta\sigma}\rangle\langle c^\dagger_{\gamma\sigma}c_{\beta\sigma}\rangle,
    \label{eq:SI_mf_decoupling}
\eal
\end{widetext}

where for simplicity we will assume a spin-restricted mean-field, i.e. $\langle c^\dagger_{\alpha\sigma}c^\dagga_{\beta\sigma}\rangle = \langle c^\dagger_{\alpha}c^\dagga_{\beta}\rangle\equiv \Delta^{mf}_{\alpha\beta}$.
Performing this mean-field approximation allows us to write the molecular Hamiltonian as

\beal
    \tilde{H}_{mol} &= \sum_I \tilde{H}^{loc}_{I} + \tilde{H}^{loc}_Y + \tilde{H}^{hyb},\\
    \tilde{H}^{loc}_{I} &= \sum_{\alpha_I\beta_I,\sigma}\tilde{t}_{\alpha_I\beta_I}\ c^\dagger_{\alpha_I\sigma}c^\dagga_{\beta_I\sigma}\\
    &+\frac{1}{2}\sum_{\substack{\alpha_I\beta_I\gamma_I\delta_I\\ \sigma\sigma'}}U_{\alpha_I\beta_I\,\gamma_I\delta_I}\ c^\dagger_{\alpha_I\sigma}c^\dagger_{\gamma_I\sigma'}c^\dagga_{\delta_I\sigma'}c^\dagga_{\beta_I\sigma},\\
    \tilde{H}^{loc}_Y &= \sum_{\alpha_Y\beta_Y,\sigma}\tilde{t}_{\alpha_Y\beta_Y}\ c^\dagger_{\alpha_Y\sigma}c^\dagga_{\beta_Y\sigma},\\
    \tilde{H}^{hyb} &= \sum_{I\neq J}\sum_{\alpha_I\beta_J,\sigma}\tilde{t}_{\alpha_I\beta_J}\ c^\dagger_{\alpha_I\sigma}c^\dagga_{\beta_J\sigma}\\
    &+\sum_{I}\sum_{\alpha_I\beta_Y,\sigma}\left(\tilde{t}_{\alpha_I\beta_Y}\ c^\dagger_{\alpha_I\sigma}c^\dagga_{\beta_Y\sigma}+\mathrm{h.c.}\right),
    \label{eq:SI_Mol_frag_mf}
\eal

were we have introduced the effective one-body terms

\beal
    \tilde{t}_{\alpha\beta} &= t_{\alpha\beta}+\sum_{I,J} u^{IJ}_{\alpha\beta} + \sum_{I}\left[u^{IY}_{\alpha\beta}+u^{YI}_{\alpha\beta}\right]+u^{YY}_{\alpha\beta}-u^{DC}_{\alpha\beta},\\
    u^{AB}_{\alpha\beta} &= \sum_{\substack{\gamma_A\in A\\ \delta_B\in B}}\left[2U_{\alpha\beta\,\gamma_A\delta_B}-U_{\alpha\delta_B\,\gamma_A\beta}\right]\Delta^{mf}_{\gamma_A\delta_B},\\
    u^{DC}_{\alpha\beta} &= \begin{cases} 
      u^{II}_{\alpha\beta} & \mathrm{if}\ \alpha\ \mathrm{and}\ \beta\ \mathrm{both\ in\ the\ same\ fragment}\ I \\
      0 & \mathrm{otherwise} 
   \end{cases}.
    \label{eq:SI_MF-decoup-Hamil}
\eal

The double counting correction $u^{DC}$ is introduced in the above, compact formula to assure that we do not decouple the interaction terms in any of the $H^{loc}_I$.

Once we have performed this mean-field decoupling, we can employ the usual ghost Gutzwiller prescription (cf. Ref.~\cite{Mejuto2023a}) for Hamiltonians with local interactions.
Essentially, the idea is obtaining the variationally optimal matrices $R^I$ and $\lambda^I$ that appear in the quasiparticle Hamiltonian in Eq.~\eqref{eq:Hqp} for each fragment.
In the infinite dimensional limit, this variational optimization can be exactly substituted by a self-consistent embedding (cf. Ref~\cite{lanata2015}) mapping the quasiparticle Hamiltonian in Eq.~\eqref{eq:Hqp} to multiple impurity models (one per correlated fragment) following

\beal
    H^{imp}_I &= H^{loc}_I + \sum_{\alpha_Ia_I}\,\Big(V^I_{\alpha_Ia_I} d^\dagger_{a_I}\, c^\dagga_{\alpha_I} + \mathrm{h.c.}\Big)\\
    &-\sum_{a_Ib_I}\, \lambda_{a_Ib_I}^{I,c}\, d^\dagger_{a_I}\,d^\dagga_{b_I}.
    \label{eq:SI_imp-H}
\eal

In these impurity models, the impurity orbitals correspond to the physical orbitals in a given correlated fragment $I$, and thus they carry with them their local Hamiltonian $H^{loc}_I$.
The quasiparticle orbitals corresponding to that same fragment (i.e. $d^\dagger_{aI}$) build the bath of the impurity model, with hybridization couplings $V^I$ and bath local potential $\lambda^{I,c}$.
These bath parameters follow directly from the $R^{I}$ and $\lambda^{I}$ of the quasiparticle Hamiltonian.
The self-consistency condition in ghost Gutzwiller then corresponds to fixing the one-body reduced density matrix of each impurity model to be equal to the local one-body reduced density matrix of the corresponding fragment in the quasiparticle Hamiltonian.
Mathematically, it reads

\begin{equation}
    \langle\, d^\dagga_{b_I}\,d^\dagger_{a_I}\,\rangle_{imp} = \delta_{a_Ib_I}-\Delta^{I,imp}_{a_Ib_I} \overset{!}{=} \Delta^{qp}_{a_Ib_I} = \langle\, d^\dagger_{a_I}\, d^\dagga_{b_I} \,\rangle_{qp}\,.
    \label{eq:SI_scf}
\end{equation}

We impose this self-consistent condition iteratively.
At the $\ell$-th iteration, we require 

\begin{equation}
    \begin{split}
        \Delta^{II,qp,\ell+1} &= \mathbb{I} - \Delta^{I,imp,\ell}_{bath-bath}, \\
        R^{I,\ell+1}\cdot\sqrt{\Delta^{II,qp,\ell+1}(\mathbb{I}-\Delta^{II,qp,\ell+1})} &= \Delta^{I,imp,\ell,t}_{bath-imp},
    \end{split}
    \label{eq:SI_scf_cond}
\end{equation}

where $\Delta^{I,imp}_{bath-bath}$ corresponds to the bath-bath block of the impurity model one-body reduced density matrix, $\Delta^{I,imp}_{bath-imp}$ denotes the bath-impurity off-diagonal block of the impurity model one-body reduced density matrix, and $\Delta^{II,qp}$ denotes the component local to fragment $I$ of the quasiparticle one-body reduced density matrix.

Operationally, a ghost Gutzwiller calculation for the Hamiltonian in Eq.~\eqref{eq:SI_MF-decoup-Hamil} starts with some guess for the $R^I$ and $\lambda^I$ matrices, and proceeds as follows:
Given the current $R^I,\lambda^I$, determine the corresponding ground state quasiparticle one-body reduced density matrix $\Delta^{IJ,qp}_{ab}=\langle d^\dagger_{aI}d^\dagga_{bJ}\rangle$.
The bath parameters for the impurity models follow then

\beal
    \sqrt{\Delta^{II,qp}\big(\mathbb{I}-\Delta^{II,qp}\big)\;}\cdot V^I = \sum_{J\neq I}\Delta^{IJ,qp}\cdot R^{I,\dagger}\cdot t_{IJ},
    \label{eq:SI_V}
\eal

and

\begin{widetext}
\beal
    \lambda^{I,c}_{a_Ib_I} = -\lambda^{I}_{a_Ib_I}+\left\{\frac{\partial}{\partial \Delta^{qp}_{a_Ib_I}}
    \left[R^I\cdot \sqrt{\Delta^{II,qp}(\mathbb{I}-\Delta^{II,qp})\;}\cdot V^I \right]+\mathrm{h.c.}\right\},
    \label{eq:SI_Lc}
\eal
\end{widetext}

where the derivative term acts only on the square-root.
Finally, the ground state one-body reduced density matrix for the impurity models defined by the $V^I, \lambda^{I,c}$ pairs is determined, and Eq.~\eqref{eq:SI_scf_cond} used to determine a new set of $R^I$ and $\Delta^{II,qp}$.
From this, a new set of $\lambda^I$ can be proposed by either fitting the $\lambda^I$ at fixed $R^I$ until the new $\Delta^{II,qp}$ are obtained (cf. Ref.~\cite{Mejuto2023a}), or Eq.~\eqref{eq:SI_Lc} can be used to obtain a new guess for $\lambda^I$ using the previous $\lambda^{I,c}$.
These second option is computationally less intensive, and typically leads to an equivalent convergence.
The iterations are repeated until the $R^I, \lambda^I$ converge within some determined threshold, in this paper 1.E-6.

The only ingredient missing is thus the mean-field $\Delta^{mf}$ in Eq.~\eqref{eq:SI_MF-decoup-Hamil}.
This can be determined self-consistently with the ghost Gutzwiller simulation by starting from some guess mean-field (e.g. restricted Hartree-Fock), solving the corresponding ghost Gutzwiller embedding, and proposing a new mean-field following the prescription (cf. Ref.~\cite{Mejuto2024})

\beal
    \Delta^{mf}_{\alpha\beta} = \begin{cases}
        \langle c^\dagger_{\alpha}c^\dagga_{\beta} \rangle_{imp} & \mathrm{if}\ \alpha, \beta\in I \\
        \sum_{ab}\,R^{I,\dagger}_{a\alpha}\,\langle d^\dagger_{a}d^\dagga_{b} \rangle_{qp}\, R^{J\dagga}_{\beta b} & \mathrm{if}\ \alpha\in I, \beta\in J, I\neq J\\
        \sum_{a}\,R^{I,\dagger}_{a\alpha}\,\langle d^\dagger_{a}d^\dagga_{\beta} \rangle_{qp}  & \mathrm{if}\ \alpha \in I, \beta\in X\\
        \langle d^\dagger_{\alpha}d^\dagga_{\beta} \rangle_{qp}  & \mathrm{if}\ \alpha, \beta\in Y\\
        \end{cases}
    \label{eq:mf_in_gGut}
\eal

Thus, we end up with a nested mean-field + ghost Gutzwiller description of correlated molecules.
This is analogous to the charge self-consistency present in DFT+DMFT approximations~\cite{Kotliar2006}.

\section{Molecular Geometries}

Here we report the geometries in $\AA$ of the $x = 0, \frac{1}{2}, 1$ points in both the H$_4$ and H$_6$ ``reactions''.
The geometries everywhere else in the $(0,1)$ interval are formed as linear interpolations of these geometries.

\begin{table}[h]
    \centering
    \begin{tabular}{|l|c|c|c|}
        \hline
         & $x = 0$ & $x = 1/2$ & $x = 1$\\
         \hline
        H$_1$ & (-2.5, 0.63) & (-0.63, 0.63) & (-0.63, 2.5) \\ 
        H$_2$ & (-2.5, -0.63) & (-0.63, -0.63) & (-0.63, -2.5) \\ 
        H$_3$ & (2.5, 0.63) & (0.63, 0.63) & (0.63, 2.5) \\ 
        H$_4$ & (2.5, -0.63) & (0.63, -0.63) & (0.63, -2.5)\\
        \hline
    \end{tabular}
    \caption{Geometries in $\AA$ the H$_4$ molecule at the $x = 0, \frac{1}{2}, 1$ points.}
    \label{tab:H4_geom}
\end{table}

\begin{table}[h]
    \centering
    \begin{tabular}{|l|c|c|c|}
        \hline
         & $x = 0$ & $x = 1/2$ & $x = 1$\\
         \hline
        H$_1$ & (0.630, 0.000) & (1.079, 0.000) & (0.375, 0.000) \\ 
        H$_2$ & (0.630, 1.260) & (0.500, $\frac{\sqrt{3}}{2}$) & (0.375, 5.000) \\ 
        H$_3$ & (-0.630, 1.260) & (-0.500, $\frac{\sqrt{3}}{2}$) & (-0.375, 5.000) \\ 
        H$_4$ & (-0.630, 0.000) & (-1.079, 0.000) & (-0.375, 0.000)\\
        H$_5$ & (-0.375, -5.000) & (-0.500, $-\frac{\sqrt{3}}{2}$) & (-0.375, -5.000) \\ 
        H$_6$ & (0.375, 5.000) & (0.500, $-\frac{\sqrt{3}}{2}$) & (0.375, -5.000) \\ 
        \hline
    \end{tabular}
    \caption{Geometries in $\AA$ of the H$_6$ molecule at the $x = 0, \frac{1}{2}, 1$ points.}
    \label{tab:H6_geom}
\end{table}

\bibliographystyle{apsrev4-2}

\begin{thebibliography}{70}%
\makeatletter
\providecommand \@ifxundefined [1]{%
 \@ifx{#1\undefined}
}%
\providecommand \@ifnum [1]{%
 \ifnum #1\expandafter \@firstoftwo
 \else \expandafter \@secondoftwo
 \fi
}%
\providecommand \@ifx [1]{%
 \ifx #1\expandafter \@firstoftwo
 \else \expandafter \@secondoftwo
 \fi
}%
\providecommand \natexlab [1]{#1}%
\providecommand \enquote  [1]{``#1''}%
\providecommand \bibnamefont  [1]{#1}%
\providecommand \bibfnamefont [1]{#1}%
\providecommand \citenamefont [1]{#1}%
\providecommand \href@noop [0]{\@secondoftwo}%
\providecommand \href [0]{\begingroup \@sanitize@url \@href}%
\providecommand \@href[1]{\@@startlink{#1}\@@href}%
\providecommand \@@href[1]{\endgroup#1\@@endlink}%
\providecommand \@sanitize@url [0]{\catcode `\\12\catcode `\$12\catcode `\&12\catcode `\#12\catcode `\^12\catcode `\_12\catcode `\%12\relax}%
\providecommand \@@startlink[1]{}%
\providecommand \@@endlink[0]{}%
\providecommand \url  [0]{\begingroup\@sanitize@url \@url }%
\providecommand \@url [1]{\endgroup\@href {#1}{\urlprefix }}%
\providecommand \urlprefix  [0]{URL }%
\providecommand \Eprint [0]{\href }%
\providecommand \doibase [0]{https://doi.org/}%
\providecommand \selectlanguage [0]{\@gobble}%
\providecommand \bibinfo  [0]{\@secondoftwo}%
\providecommand \bibfield  [0]{\@secondoftwo}%
\providecommand \translation [1]{[#1]}%
\providecommand \BibitemOpen [0]{}%
\providecommand \bibitemStop [0]{}%
\providecommand \bibitemNoStop [0]{.\EOS\space}%
\providecommand \EOS [0]{\spacefactor3000\relax}%
\providecommand \BibitemShut  [1]{\csname bibitem#1\endcsname}%
\let\auto@bib@innerbib\@empty
\bibitem [{\citenamefont {H\"uckel}(1931{\natexlab{a}})}]{Huckel1931a}%
  \BibitemOpen
  \bibfield  {author} {\bibinfo {author} {\bibfnamefont {E.}~\bibnamefont {H\"uckel}},\ }\href@noop {} {\bibfield  {journal} {\bibinfo  {journal} {Z. Physik}\ }\textbf {\bibinfo {volume} {70}},\ \bibinfo {pages} {204} (\bibinfo {year} {1931}{\natexlab{a}})}\BibitemShut {NoStop}%
\bibitem [{\citenamefont {H\"uckel}(1931{\natexlab{b}})}]{Huckel1931b}%
  \BibitemOpen
  \bibfield  {author} {\bibinfo {author} {\bibfnamefont {E.}~\bibnamefont {H\"uckel}},\ }\href@noop {} {\bibfield  {journal} {\bibinfo  {journal} {Z. Physik}\ }\textbf {\bibinfo {volume} {72}},\ \bibinfo {pages} {310} (\bibinfo {year} {1931}{\natexlab{b}})}\BibitemShut {NoStop}%
\bibitem [{\citenamefont {H\"uckel}(1932)}]{Huckel1932}%
  \BibitemOpen
  \bibfield  {author} {\bibinfo {author} {\bibfnamefont {E.}~\bibnamefont {H\"uckel}},\ }\href@noop {} {\bibfield  {journal} {\bibinfo  {journal} {Z. Physik}\ }\textbf {\bibinfo {volume} {76}},\ \bibinfo {pages} {628} (\bibinfo {year} {1932})}\BibitemShut {NoStop}%
\bibitem [{\citenamefont {Goodenough}(1955)}]{Goodenough1955}%
  \BibitemOpen
  \bibfield  {author} {\bibinfo {author} {\bibfnamefont {J.~B.}\ \bibnamefont {Goodenough}},\ }\href@noop {} {\bibfield  {journal} {\bibinfo  {journal} {Phys. Rev.}\ }\textbf {\bibinfo {volume} {100}},\ \bibinfo {pages} {564} (\bibinfo {year} {1955})}\BibitemShut {NoStop}%
\bibitem [{\citenamefont {Kanamori}(1959)}]{KANAMORI1959}%
  \BibitemOpen
  \bibfield  {author} {\bibinfo {author} {\bibfnamefont {J.}~\bibnamefont {Kanamori}},\ }\href@noop {} {\bibfield  {journal} {\bibinfo  {journal} {Journal of Physics and Chemistry of Solids}\ }\textbf {\bibinfo {volume} {10}},\ \bibinfo {pages} {87} (\bibinfo {year} {1959})}\BibitemShut {NoStop}%
\bibitem [{\citenamefont {Woodward}\ and\ \citenamefont {Hoffmann}(1965)}]{Woodward1965}%
  \BibitemOpen
  \bibfield  {author} {\bibinfo {author} {\bibfnamefont {R.~B.}\ \bibnamefont {Woodward}}\ and\ \bibinfo {author} {\bibfnamefont {R.}~\bibnamefont {Hoffmann}},\ }\href {https://doi.org/10.1021/ja01080a054} {\bibfield  {journal} {\bibinfo  {journal} {Journal of the American Chemical Society}\ }\textbf {\bibinfo {volume} {87}},\ \bibinfo {pages} {395} (\bibinfo {year} {1965})},\ \Eprint {https://arxiv.org/abs/https://doi.org/10.1021/ja01080a054} {https://doi.org/10.1021/ja01080a054} \BibitemShut {NoStop}%
\bibitem [{\citenamefont {Longuet-Higgins}\ and\ \citenamefont {Abrahamson}(1965)}]{Longuet-Higgins1965}%
  \BibitemOpen
  \bibfield  {author} {\bibinfo {author} {\bibfnamefont {H.~C.}\ \bibnamefont {Longuet-Higgins}}\ and\ \bibinfo {author} {\bibfnamefont {E.~W.}\ \bibnamefont {Abrahamson}},\ }\href@noop {} {\bibfield  {journal} {\bibinfo  {journal} {Journal of the American Chemical Society}\ }\textbf {\bibinfo {volume} {87}},\ \bibinfo {pages} {2045} (\bibinfo {year} {1965})}\BibitemShut {NoStop}%
\bibitem [{\citenamefont {Zimmerman}(1966)}]{Zimmerman1966}%
  \BibitemOpen
  \bibfield  {author} {\bibinfo {author} {\bibfnamefont {H.~E.}\ \bibnamefont {Zimmerman}},\ }\href@noop {} {\bibfield  {journal} {\bibinfo  {journal} {Journal of the American Chemical Society}\ }\textbf {\bibinfo {volume} {88}},\ \bibinfo {pages} {1564} (\bibinfo {year} {1966})}\BibitemShut {NoStop}%
\bibitem [{\citenamefont {Woodward}\ and\ \citenamefont {Hoffmann}(1969)}]{Woodward1969}%
  \BibitemOpen
  \bibfield  {author} {\bibinfo {author} {\bibfnamefont {R.~B.}\ \bibnamefont {Woodward}}\ and\ \bibinfo {author} {\bibfnamefont {R.}~\bibnamefont {Hoffmann}},\ }\href@noop {} {\bibfield  {journal} {\bibinfo  {journal} {Angewandte Chemie International Edition in English}\ }\textbf {\bibinfo {volume} {8}},\ \bibinfo {pages} {781} (\bibinfo {year} {1969})}\BibitemShut {NoStop}%
\bibitem [{\citenamefont {Dewar}(1971)}]{Dewar1971}%
  \BibitemOpen
  \bibfield  {author} {\bibinfo {author} {\bibfnamefont {M.~J.~S.}\ \bibnamefont {Dewar}},\ }\href@noop {} {\bibfield  {journal} {\bibinfo  {journal} {Angewandte Chemie International Edition in English}\ }\textbf {\bibinfo {volume} {10}},\ \bibinfo {pages} {761} (\bibinfo {year} {1971})}\BibitemShut {NoStop}%
\bibitem [{\citenamefont {Fukui}(1982)}]{Fukui1982}%
  \BibitemOpen
  \bibfield  {author} {\bibinfo {author} {\bibfnamefont {K.}~\bibnamefont {Fukui}},\ }\href@noop {} {\bibfield  {journal} {\bibinfo  {journal} {Science}\ }\textbf {\bibinfo {volume} {218}},\ \bibinfo {pages} {747} (\bibinfo {year} {1982})}\BibitemShut {NoStop}%
\bibitem [{\citenamefont {Xie}\ \emph {et~al.}(2025)\citenamefont {Xie}, \citenamefont {Mirzanejad},\ and\ \citenamefont {Muechler}}]{Xie2025}%
  \BibitemOpen
  \bibfield  {author} {\bibinfo {author} {\bibfnamefont {Z.}~\bibnamefont {Xie}}, \bibinfo {author} {\bibfnamefont {A.}~\bibnamefont {Mirzanejad}},\ and\ \bibinfo {author} {\bibfnamefont {L.}~\bibnamefont {Muechler}},\ }\href {https://arxiv.org/abs/2506.18984} {\bibinfo {title} {Topological transitions in orbital-symmetry-controlled chemical reactions}} (\bibinfo {year} {2025}),\ \Eprint {https://arxiv.org/abs/2506.18984} {arXiv:2506.18984 [cond-mat.str-el]} \BibitemShut {NoStop}%
\bibitem [{\citenamefont {Fabrizio}(2022)}]{mio-2}%
  \BibitemOpen
  \bibfield  {author} {\bibinfo {author} {\bibfnamefont {M.}~\bibnamefont {Fabrizio}},\ }\href {https://doi.org/10.1038/s41467-022-29190-y} {\bibfield  {journal} {\bibinfo  {journal} {Nature Communications}\ }\textbf {\bibinfo {volume} {13}},\ \bibinfo {pages} {1561} (\bibinfo {year} {2022})}\BibitemShut {NoStop}%
\bibitem [{\citenamefont {Fabrizio}(2023)}]{mio-Mott}%
  \BibitemOpen
  \bibfield  {author} {\bibinfo {author} {\bibfnamefont {M.}~\bibnamefont {Fabrizio}},\ }\href {https://doi.org/10.1103/PhysRevLett.130.156702} {\bibfield  {journal} {\bibinfo  {journal} {Phys. Rev. Lett.}\ }\textbf {\bibinfo {volume} {130}},\ \bibinfo {pages} {156702} (\bibinfo {year} {2023})}\BibitemShut {NoStop}%
\bibitem [{\citenamefont {Martin}(2004)}]{martin2004}%
  \BibitemOpen
  \bibfield  {author} {\bibinfo {author} {\bibfnamefont {R.~M.}\ \bibnamefont {Martin}},\ }\href {https://doi.org/10.1017/CBO9780511805769} {\emph {\bibinfo {title} {Electronic Structure: Basic Theory and Practical Methods}}}\ (\bibinfo  {publisher} {Cambridge University Press},\ \bibinfo {year} {2004})\BibitemShut {NoStop}%
\bibitem [{\citenamefont {Lanat\`a}\ \emph {et~al.}(2017)\citenamefont {Lanat\`a}, \citenamefont {Lee}, \citenamefont {Yao},\ and\ \citenamefont {Dobrosavljevi\ifmmode~\acute{c}\else \'{c}\fi{}}}]{Lanata2017}%
  \BibitemOpen
  \bibfield  {author} {\bibinfo {author} {\bibfnamefont {N.}~\bibnamefont {Lanat\`a}}, \bibinfo {author} {\bibfnamefont {T.-H.}\ \bibnamefont {Lee}}, \bibinfo {author} {\bibfnamefont {Y.-X.}\ \bibnamefont {Yao}},\ and\ \bibinfo {author} {\bibfnamefont {V.}~\bibnamefont {Dobrosavljevi\ifmmode~\acute{c}\else \'{c}\fi{}}},\ }\href {https://doi.org/10.1103/PhysRevB.96.195126} {\bibfield  {journal} {\bibinfo  {journal} {Phys. Rev. B}\ }\textbf {\bibinfo {volume} {96}},\ \bibinfo {pages} {195126} (\bibinfo {year} {2017})}\BibitemShut {NoStop}%
\bibitem [{\citenamefont {Guerci}\ \emph {et~al.}(2019)\citenamefont {Guerci}, \citenamefont {Capone},\ and\ \citenamefont {Fabrizio}}]{guerci2019}%
  \BibitemOpen
  \bibfield  {author} {\bibinfo {author} {\bibfnamefont {D.}~\bibnamefont {Guerci}}, \bibinfo {author} {\bibfnamefont {M.}~\bibnamefont {Capone}},\ and\ \bibinfo {author} {\bibfnamefont {M.}~\bibnamefont {Fabrizio}},\ }\href@noop {} {\bibfield  {journal} {\bibinfo  {journal} {Phys. Rev. Mater.}\ }\textbf {\bibinfo {volume} {3}},\ \bibinfo {pages} {054605} (\bibinfo {year} {2019})}\BibitemShut {NoStop}%
\bibitem [{\citenamefont {Frank}\ \emph {et~al.}(2021)\citenamefont {Frank}, \citenamefont {Lee}, \citenamefont {Bhattacharyya}, \citenamefont {Tsang}, \citenamefont {Quito}, \citenamefont {Dobrosavljevi{\'c}}, \citenamefont {Christiansen},\ and\ \citenamefont {Lanat{\`a}}}]{frank2021}%
  \BibitemOpen
  \bibfield  {author} {\bibinfo {author} {\bibfnamefont {M.~S.}\ \bibnamefont {Frank}}, \bibinfo {author} {\bibfnamefont {T.-H.}\ \bibnamefont {Lee}}, \bibinfo {author} {\bibfnamefont {G.}~\bibnamefont {Bhattacharyya}}, \bibinfo {author} {\bibfnamefont {P.~K.~H.}\ \bibnamefont {Tsang}}, \bibinfo {author} {\bibfnamefont {V.~L.}\ \bibnamefont {Quito}}, \bibinfo {author} {\bibfnamefont {V.}~\bibnamefont {Dobrosavljevi{\'c}}}, \bibinfo {author} {\bibfnamefont {O.}~\bibnamefont {Christiansen}},\ and\ \bibinfo {author} {\bibfnamefont {N.}~\bibnamefont {Lanat{\`a}}},\ }\href@noop {} {\bibfield  {journal} {\bibinfo  {journal} {Phys. Rev. B}\ }\textbf {\bibinfo {volume} {104}},\ \bibinfo {pages} {L081103} (\bibinfo {year} {2021})}\BibitemShut {NoStop}%
\bibitem [{\citenamefont {Mejuto-Zaera}\ and\ \citenamefont {Fabrizio}(2023)}]{Mejuto2023a}%
  \BibitemOpen
  \bibfield  {author} {\bibinfo {author} {\bibfnamefont {C.}~\bibnamefont {Mejuto-Zaera}}\ and\ \bibinfo {author} {\bibfnamefont {M.}~\bibnamefont {Fabrizio}},\ }\href {https://doi.org/10.1103/PhysRevB.107.235150} {\bibfield  {journal} {\bibinfo  {journal} {Phys. Rev. B}\ }\textbf {\bibinfo {volume} {107}},\ \bibinfo {pages} {235150} (\bibinfo {year} {2023})}\BibitemShut {NoStop}%
\bibitem [{\citenamefont {Lee}\ \emph {et~al.}(2023{\natexlab{a}})\citenamefont {Lee}, \citenamefont {Melnick}, \citenamefont {Adler}, \citenamefont {Lanat\`a},\ and\ \citenamefont {Kotliar}}]{Lee2023a}%
  \BibitemOpen
  \bibfield  {author} {\bibinfo {author} {\bibfnamefont {T.-H.}\ \bibnamefont {Lee}}, \bibinfo {author} {\bibfnamefont {C.}~\bibnamefont {Melnick}}, \bibinfo {author} {\bibfnamefont {R.}~\bibnamefont {Adler}}, \bibinfo {author} {\bibfnamefont {N.}~\bibnamefont {Lanat\`a}},\ and\ \bibinfo {author} {\bibfnamefont {G.}~\bibnamefont {Kotliar}},\ }\href {https://doi.org/10.1103/PhysRevB.108.245147} {\bibfield  {journal} {\bibinfo  {journal} {Phys. Rev. B}\ }\textbf {\bibinfo {volume} {108}},\ \bibinfo {pages} {245147} (\bibinfo {year} {2023}{\natexlab{a}})}\BibitemShut {NoStop}%
\bibitem [{\citenamefont {Lee}\ \emph {et~al.}(2023{\natexlab{b}})\citenamefont {Lee}, \citenamefont {Lanat\`a},\ and\ \citenamefont {Kotliar}}]{Lee2023b}%
  \BibitemOpen
  \bibfield  {author} {\bibinfo {author} {\bibfnamefont {T.-H.}\ \bibnamefont {Lee}}, \bibinfo {author} {\bibfnamefont {N.}~\bibnamefont {Lanat\`a}},\ and\ \bibinfo {author} {\bibfnamefont {G.}~\bibnamefont {Kotliar}},\ }\href {https://doi.org/10.1103/PhysRevB.107.L121104} {\bibfield  {journal} {\bibinfo  {journal} {Phys. Rev. B}\ }\textbf {\bibinfo {volume} {107}},\ \bibinfo {pages} {L121104} (\bibinfo {year} {2023}{\natexlab{b}})}\BibitemShut {NoStop}%
\bibitem [{\citenamefont {Guerci}\ \emph {et~al.}(2023)\citenamefont {Guerci}, \citenamefont {Capone},\ and\ \citenamefont {Lanat\`a}}]{Guerci2023}%
  \BibitemOpen
  \bibfield  {author} {\bibinfo {author} {\bibfnamefont {D.}~\bibnamefont {Guerci}}, \bibinfo {author} {\bibfnamefont {M.}~\bibnamefont {Capone}},\ and\ \bibinfo {author} {\bibfnamefont {N.}~\bibnamefont {Lanat\`a}},\ }\href {https://doi.org/10.1103/PhysRevResearch.5.L032023} {\bibfield  {journal} {\bibinfo  {journal} {Phys. Rev. Res.}\ }\textbf {\bibinfo {volume} {5}},\ \bibinfo {pages} {L032023} (\bibinfo {year} {2023})}\BibitemShut {NoStop}%
\bibitem [{\citenamefont {Blason}\ and\ \citenamefont {Fabrizio}(2023)}]{Andrea-PRB2023}%
  \BibitemOpen
  \bibfield  {author} {\bibinfo {author} {\bibfnamefont {A.}~\bibnamefont {Blason}}\ and\ \bibinfo {author} {\bibfnamefont {M.}~\bibnamefont {Fabrizio}},\ }\href {https://doi.org/10.1103/PhysRevB.108.125115} {\bibfield  {journal} {\bibinfo  {journal} {Phys. Rev. B}\ }\textbf {\bibinfo {volume} {108}},\ \bibinfo {pages} {125115} (\bibinfo {year} {2023})}\BibitemShut {NoStop}%
\bibitem [{\citenamefont {Pasqua}\ and\ \citenamefont {Fabrizio}(2025)}]{Ivan-SciPost}%
  \BibitemOpen
  \bibfield  {author} {\bibinfo {author} {\bibfnamefont {I.}~\bibnamefont {Pasqua}}\ and\ \bibinfo {author} {\bibfnamefont {M.}~\bibnamefont {Fabrizio}},\ }\href {https://doi.org/10.21468/SciPostPhys.19.1.014} {\bibfield  {journal} {\bibinfo  {journal} {SciPost Phys.}\ }\textbf {\bibinfo {volume} {19}},\ \bibinfo {pages} {014} (\bibinfo {year} {2025})}\BibitemShut {NoStop}%
\bibitem [{\citenamefont {Skolimowski}\ and\ \citenamefont {Fabrizio}(2022)}]{Jan}%
  \BibitemOpen
  \bibfield  {author} {\bibinfo {author} {\bibfnamefont {J.}~\bibnamefont {Skolimowski}}\ and\ \bibinfo {author} {\bibfnamefont {M.}~\bibnamefont {Fabrizio}},\ }\href {https://doi.org/10.1103/PhysRevB.106.045109} {\bibfield  {journal} {\bibinfo  {journal} {Phys. Rev. B}\ }\textbf {\bibinfo {volume} {106}},\ \bibinfo {pages} {045109} (\bibinfo {year} {2022})}\BibitemShut {NoStop}%
\bibitem [{\citenamefont {Luttinger}(1960)}]{Luttinger}%
  \BibitemOpen
  \bibfield  {author} {\bibinfo {author} {\bibfnamefont {J.~M.}\ \bibnamefont {Luttinger}},\ }\href {https://doi.org/10.1103/PhysRev.119.1153} {\bibfield  {journal} {\bibinfo  {journal} {Phys. Rev.}\ }\textbf {\bibinfo {volume} {119}},\ \bibinfo {pages} {1153} (\bibinfo {year} {1960})}\BibitemShut {NoStop}%
\bibitem [{\citenamefont {Gutzwiller}(1963)}]{Gutzwiller1963}%
  \BibitemOpen
  \bibfield  {author} {\bibinfo {author} {\bibfnamefont {M.~C.}\ \bibnamefont {Gutzwiller}},\ }\href {https://doi.org/10.1103/PhysRevLett.10.159} {\bibfield  {journal} {\bibinfo  {journal} {Phys. Rev. Lett.}\ }\textbf {\bibinfo {volume} {10}},\ \bibinfo {pages} {159} (\bibinfo {year} {1963})}\BibitemShut {NoStop}%
\bibitem [{\citenamefont {Gutzwiller}(1965)}]{Gutzwiller1965}%
  \BibitemOpen
  \bibfield  {author} {\bibinfo {author} {\bibfnamefont {M.~C.}\ \bibnamefont {Gutzwiller}},\ }\href {https://doi.org/10.1103/PhysRev.137.A1726} {\bibfield  {journal} {\bibinfo  {journal} {Phys. Rev.}\ }\textbf {\bibinfo {volume} {137}},\ \bibinfo {pages} {A1726} (\bibinfo {year} {1965})}\BibitemShut {NoStop}%
\bibitem [{\citenamefont {B\"unemann}\ \emph {et~al.}(1998)\citenamefont {B\"unemann}, \citenamefont {Weber},\ and\ \citenamefont {Gebhard}}]{Bunemann1998}%
  \BibitemOpen
  \bibfield  {author} {\bibinfo {author} {\bibfnamefont {J.}~\bibnamefont {B\"unemann}}, \bibinfo {author} {\bibfnamefont {W.}~\bibnamefont {Weber}},\ and\ \bibinfo {author} {\bibfnamefont {F.}~\bibnamefont {Gebhard}},\ }\href@noop {} {\bibfield  {journal} {\bibinfo  {journal} {Phys. Rev. B}\ }\textbf {\bibinfo {volume} {57}},\ \bibinfo {pages} {6896} (\bibinfo {year} {1998})}\BibitemShut {NoStop}%
\bibitem [{\citenamefont {Fabrizio}(2007)}]{Fabrizio2007}%
  \BibitemOpen
  \bibfield  {author} {\bibinfo {author} {\bibfnamefont {M.}~\bibnamefont {Fabrizio}},\ }\href {https://doi.org/10.1103/PhysRevB.76.165110} {\bibfield  {journal} {\bibinfo  {journal} {Phys. Rev. B}\ }\textbf {\bibinfo {volume} {76}},\ \bibinfo {pages} {165110} (\bibinfo {year} {2007})}\BibitemShut {NoStop}%
\bibitem [{\citenamefont {Yao}\ \emph {et~al.}(2014)\citenamefont {Yao}, \citenamefont {Liu}, \citenamefont {Wang},\ and\ \citenamefont {Ho}}]{Yao2014}%
  \BibitemOpen
  \bibfield  {author} {\bibinfo {author} {\bibfnamefont {Y.~X.}\ \bibnamefont {Yao}}, \bibinfo {author} {\bibfnamefont {J.}~\bibnamefont {Liu}}, \bibinfo {author} {\bibfnamefont {C.~Z.}\ \bibnamefont {Wang}},\ and\ \bibinfo {author} {\bibfnamefont {K.~M.}\ \bibnamefont {Ho}},\ }\href {https://doi.org/10.1103/PhysRevB.89.045131} {\bibfield  {journal} {\bibinfo  {journal} {Phys. Rev. B}\ }\textbf {\bibinfo {volume} {89}},\ \bibinfo {pages} {045131} (\bibinfo {year} {2014})}\BibitemShut {NoStop}%
\bibitem [{\citenamefont {Yao}\ \emph {et~al.}(2015)\citenamefont {Yao}, \citenamefont {Liu}, \citenamefont {Liu}, \citenamefont {Lu}, \citenamefont {Wang},\ and\ \citenamefont {Ho}}]{Yao2015}%
  \BibitemOpen
  \bibfield  {author} {\bibinfo {author} {\bibfnamefont {Y.~X.}\ \bibnamefont {Yao}}, \bibinfo {author} {\bibfnamefont {J.}~\bibnamefont {Liu}}, \bibinfo {author} {\bibfnamefont {C.}~\bibnamefont {Liu}}, \bibinfo {author} {\bibfnamefont {W.~C.}\ \bibnamefont {Lu}}, \bibinfo {author} {\bibfnamefont {C.~Z.}\ \bibnamefont {Wang}},\ and\ \bibinfo {author} {\bibfnamefont {K.~M.}\ \bibnamefont {Ho}},\ }\href {https://www.nature.com/articles/srep13478} {\bibfield  {journal} {\bibinfo  {journal} {Scientific Reports}\ }\textbf {\bibinfo {volume} {5}},\ \bibinfo {pages} {13478} (\bibinfo {year} {2015})}\BibitemShut {NoStop}%
\bibitem [{\citenamefont {Lanat{\`a}}\ \emph {et~al.}(2015)\citenamefont {Lanat{\`a}}, \citenamefont {Yao}, \citenamefont {Wang}, \citenamefont {Ho},\ and\ \citenamefont {Kotliar}}]{lanata2015}%
  \BibitemOpen
  \bibfield  {author} {\bibinfo {author} {\bibfnamefont {N.}~\bibnamefont {Lanat{\`a}}}, \bibinfo {author} {\bibfnamefont {Y.}~\bibnamefont {Yao}}, \bibinfo {author} {\bibfnamefont {C.-Z.}\ \bibnamefont {Wang}}, \bibinfo {author} {\bibfnamefont {K.-M.}\ \bibnamefont {Ho}},\ and\ \bibinfo {author} {\bibfnamefont {G.}~\bibnamefont {Kotliar}},\ }\href@noop {} {\bibfield  {journal} {\bibinfo  {journal} {Phys. Rev. X}\ }\textbf {\bibinfo {volume} {5}},\ \bibinfo {pages} {011008} (\bibinfo {year} {2015})}\BibitemShut {NoStop}%
\bibitem [{\citenamefont {Fabrizio}(2017)}]{fabrizio2017}%
  \BibitemOpen
  \bibfield  {author} {\bibinfo {author} {\bibfnamefont {M.}~\bibnamefont {Fabrizio}},\ }\href@noop {} {\bibfield  {journal} {\bibinfo  {journal} {Phys. Rev. B}\ }\textbf {\bibinfo {volume} {95}},\ \bibinfo {pages} {075156} (\bibinfo {year} {2017})}\BibitemShut {NoStop}%
\bibitem [{\citenamefont {Georges}\ \emph {et~al.}(1996)\citenamefont {Georges}, \citenamefont {Kotliar}, \citenamefont {Krauth},\ and\ \citenamefont {Rozenberg}}]{Kotliar1996}%
  \BibitemOpen
  \bibfield  {author} {\bibinfo {author} {\bibfnamefont {A.}~\bibnamefont {Georges}}, \bibinfo {author} {\bibfnamefont {G.}~\bibnamefont {Kotliar}}, \bibinfo {author} {\bibfnamefont {W.}~\bibnamefont {Krauth}},\ and\ \bibinfo {author} {\bibfnamefont {M.~J.}\ \bibnamefont {Rozenberg}},\ }\href@noop {} {\bibfield  {journal} {\bibinfo  {journal} {Rev. Mod. Phys.}\ }\textbf {\bibinfo {volume} {68}},\ \bibinfo {pages} {13} (\bibinfo {year} {1996})}\BibitemShut {NoStop}%
\bibitem [{\citenamefont {Kotliar}\ \emph {et~al.}(2006)\citenamefont {Kotliar}, \citenamefont {Savrasov}, \citenamefont {Haule}, \citenamefont {Oudovenko}, \citenamefont {Parcollet},\ and\ \citenamefont {Marianetti}}]{Kotliar2006}%
  \BibitemOpen
  \bibfield  {author} {\bibinfo {author} {\bibfnamefont {G.}~\bibnamefont {Kotliar}}, \bibinfo {author} {\bibfnamefont {S.~Y.}\ \bibnamefont {Savrasov}}, \bibinfo {author} {\bibfnamefont {K.}~\bibnamefont {Haule}}, \bibinfo {author} {\bibfnamefont {V.~S.}\ \bibnamefont {Oudovenko}}, \bibinfo {author} {\bibfnamefont {O.}~\bibnamefont {Parcollet}},\ and\ \bibinfo {author} {\bibfnamefont {C.}~\bibnamefont {Marianetti}},\ }\href@noop {} {\bibfield  {journal} {\bibinfo  {journal} {Rev. Mod. Phys.}\ }\textbf {\bibinfo {volume} {78}},\ \bibinfo {pages} {865} (\bibinfo {year} {2006})}\BibitemShut {NoStop}%
\bibitem [{\citenamefont {Kotliar}\ and\ \citenamefont {Abrahams}(2001)}]{Kotliar2001b}%
  \BibitemOpen
  \bibfield  {author} {\bibinfo {author} {\bibfnamefont {G.}~\bibnamefont {Kotliar}}\ and\ \bibinfo {author} {\bibfnamefont {E.}~\bibnamefont {Abrahams}},\ }\href@noop {} {\bibfield  {journal} {\bibinfo  {journal} {Nature}\ }\textbf {\bibinfo {volume} {410}},\ \bibinfo {pages} {793} (\bibinfo {year} {2001})}\BibitemShut {NoStop}%
\bibitem [{\citenamefont {Arita}\ \emph {et~al.}(2007)\citenamefont {Arita}, \citenamefont {Held}, \citenamefont {Lukoyanov},\ and\ \citenamefont {Anisimov}}]{Arita2007}%
  \BibitemOpen
  \bibfield  {author} {\bibinfo {author} {\bibfnamefont {R.}~\bibnamefont {Arita}}, \bibinfo {author} {\bibfnamefont {K.}~\bibnamefont {Held}}, \bibinfo {author} {\bibfnamefont {A.~V.}\ \bibnamefont {Lukoyanov}},\ and\ \bibinfo {author} {\bibfnamefont {V.~I.}\ \bibnamefont {Anisimov}},\ }\href@noop {} {\bibfield  {journal} {\bibinfo  {journal} {Phys. Rev. Lett.}\ }\textbf {\bibinfo {volume} {98}},\ \bibinfo {pages} {166402} (\bibinfo {year} {2007})}\BibitemShut {NoStop}%
\bibitem [{\citenamefont {Shim}\ \emph {et~al.}(2007)\citenamefont {Shim}, \citenamefont {Haule},\ and\ \citenamefont {Kotliar}}]{Shim2007}%
  \BibitemOpen
  \bibfield  {author} {\bibinfo {author} {\bibfnamefont {J.~H.}\ \bibnamefont {Shim}}, \bibinfo {author} {\bibfnamefont {K.}~\bibnamefont {Haule}},\ and\ \bibinfo {author} {\bibfnamefont {G.}~\bibnamefont {Kotliar}},\ }\href {https://doi.org/10.1126/science.1149064} {\bibfield  {journal} {\bibinfo  {journal} {Science}\ }\textbf {\bibinfo {volume} {318}},\ \bibinfo {pages} {1615} (\bibinfo {year} {2007})}\BibitemShut {NoStop}%
\bibitem [{\citenamefont {Takizawa}\ \emph {et~al.}(2009)\citenamefont {Takizawa}, \citenamefont {Minohara}, \citenamefont {Kumigashira}, \citenamefont {Toyota}, \citenamefont {Oshima}, \citenamefont {Wadati}, \citenamefont {Yoshida}, \citenamefont {Fujimori}, \citenamefont {Lippmaa}, \citenamefont {Kawasaki}, \citenamefont {Koinuma}, \citenamefont {Sordi},\ and\ \citenamefont {Rozenberg}}]{Takizawa2009}%
  \BibitemOpen
  \bibfield  {author} {\bibinfo {author} {\bibfnamefont {M.}~\bibnamefont {Takizawa}}, \bibinfo {author} {\bibfnamefont {M.}~\bibnamefont {Minohara}}, \bibinfo {author} {\bibfnamefont {H.}~\bibnamefont {Kumigashira}}, \bibinfo {author} {\bibfnamefont {D.}~\bibnamefont {Toyota}}, \bibinfo {author} {\bibfnamefont {M.}~\bibnamefont {Oshima}}, \bibinfo {author} {\bibfnamefont {H.}~\bibnamefont {Wadati}}, \bibinfo {author} {\bibfnamefont {T.}~\bibnamefont {Yoshida}}, \bibinfo {author} {\bibfnamefont {A.}~\bibnamefont {Fujimori}}, \bibinfo {author} {\bibfnamefont {M.}~\bibnamefont {Lippmaa}}, \bibinfo {author} {\bibfnamefont {M.}~\bibnamefont {Kawasaki}}, \bibinfo {author} {\bibfnamefont {H.}~\bibnamefont {Koinuma}}, \bibinfo {author} {\bibfnamefont {G.}~\bibnamefont {Sordi}},\ and\ \bibinfo {author} {\bibfnamefont {M.}~\bibnamefont {Rozenberg}},\ }\href@noop {} {\bibfield  {journal} {\bibinfo  {journal} {Phys. Rev. B}\ }\textbf {\bibinfo {volume} {80}},\ \bibinfo {pages} {235104} (\bibinfo {year}
  {2009})}\BibitemShut {NoStop}%
\bibitem [{\citenamefont {Haule}\ \emph {et~al.}(2010)\citenamefont {Haule}, \citenamefont {Yee},\ and\ \citenamefont {Kim}}]{Haule2010}%
  \BibitemOpen
  \bibfield  {author} {\bibinfo {author} {\bibfnamefont {K.}~\bibnamefont {Haule}}, \bibinfo {author} {\bibfnamefont {C.-H.}\ \bibnamefont {Yee}},\ and\ \bibinfo {author} {\bibfnamefont {K.}~\bibnamefont {Kim}},\ }\href@noop {} {\bibfield  {journal} {\bibinfo  {journal} {Phys. Rev. B}\ }\textbf {\bibinfo {volume} {81}},\ \bibinfo {pages} {195107} (\bibinfo {year} {2010})}\BibitemShut {NoStop}%
\bibitem [{\citenamefont {Park}\ \emph {et~al.}(2014)\citenamefont {Park}, \citenamefont {Millis},\ and\ \citenamefont {Marianetti}}]{Park2014}%
  \BibitemOpen
  \bibfield  {author} {\bibinfo {author} {\bibfnamefont {H.}~\bibnamefont {Park}}, \bibinfo {author} {\bibfnamefont {A.~J.}\ \bibnamefont {Millis}},\ and\ \bibinfo {author} {\bibfnamefont {C.~A.}\ \bibnamefont {Marianetti}},\ }\href@noop {} {\bibfield  {journal} {\bibinfo  {journal} {Phys. Rev. B}\ }\textbf {\bibinfo {volume} {90}},\ \bibinfo {pages} {235103} (\bibinfo {year} {2014})}\BibitemShut {NoStop}%
\bibitem [{\citenamefont {Haule}\ and\ \citenamefont {Birol}(2015)}]{Haule2015}%
  \BibitemOpen
  \bibfield  {author} {\bibinfo {author} {\bibfnamefont {K.}~\bibnamefont {Haule}}\ and\ \bibinfo {author} {\bibfnamefont {T.}~\bibnamefont {Birol}},\ }\href@noop {} {\bibfield  {journal} {\bibinfo  {journal} {Phys. Rev. Lett.}\ }\textbf {\bibinfo {volume} {115}},\ \bibinfo {pages} {256402} (\bibinfo {year} {2015})}\BibitemShut {NoStop}%
\bibitem [{\citenamefont {Paul}\ and\ \citenamefont {Birol}(2019)}]{Paul2019}%
  \BibitemOpen
  \bibfield  {author} {\bibinfo {author} {\bibfnamefont {A.}~\bibnamefont {Paul}}\ and\ \bibinfo {author} {\bibfnamefont {T.}~\bibnamefont {Birol}},\ }\href@noop {} {\bibfield  {journal} {\bibinfo  {journal} {Annu. Rev. Mater. Res.}\ }\textbf {\bibinfo {volume} {49}},\ \bibinfo {pages} {31} (\bibinfo {year} {2019})}\BibitemShut {NoStop}%
\bibitem [{\citenamefont {Zhu}\ \emph {et~al.}(2020)\citenamefont {Zhu}, \citenamefont {Cui},\ and\ \citenamefont {Chan}}]{Zhu2020}%
  \BibitemOpen
  \bibfield  {author} {\bibinfo {author} {\bibfnamefont {T.}~\bibnamefont {Zhu}}, \bibinfo {author} {\bibfnamefont {Z.-H.}\ \bibnamefont {Cui}},\ and\ \bibinfo {author} {\bibfnamefont {G.~K.-L.}\ \bibnamefont {Chan}},\ }\href {https://doi.org/10.1021/acs.jctc.9b00934} {\bibfield  {journal} {\bibinfo  {journal} {J. Chem. Theory Comput.}\ }\textbf {\bibinfo {volume} {16}},\ \bibinfo {pages} {141} (\bibinfo {year} {2020})},\ \bibinfo {note} {pMID: 31815457}\BibitemShut {NoStop}%
\bibitem [{\citenamefont {Zhu}\ and\ \citenamefont {Chan}(2021)}]{Zhu2021}%
  \BibitemOpen
  \bibfield  {author} {\bibinfo {author} {\bibfnamefont {T.}~\bibnamefont {Zhu}}\ and\ \bibinfo {author} {\bibfnamefont {G.~K.-L.}\ \bibnamefont {Chan}},\ }\href {https://doi.org/10.1103/PhysRevX.11.021006} {\bibfield  {journal} {\bibinfo  {journal} {Phys. Rev. X}\ }\textbf {\bibinfo {volume} {11}},\ \bibinfo {pages} {021006} (\bibinfo {year} {2021})}\BibitemShut {NoStop}%
\bibitem [{\citenamefont {Knizia}\ and\ \citenamefont {Chan}(2012)}]{Knizia2012}%
  \BibitemOpen
  \bibfield  {author} {\bibinfo {author} {\bibfnamefont {G.}~\bibnamefont {Knizia}}\ and\ \bibinfo {author} {\bibfnamefont {G.~K.-L.}\ \bibnamefont {Chan}},\ }\href {https://doi.org/10.1103/PhysRevLett.109.186404} {\bibfield  {journal} {\bibinfo  {journal} {Phys. Rev. Lett.}\ }\textbf {\bibinfo {volume} {109}},\ \bibinfo {pages} {186404} (\bibinfo {year} {2012})}\BibitemShut {NoStop}%
\bibitem [{\citenamefont {Wouters}\ \emph {et~al.}(2016)\citenamefont {Wouters}, \citenamefont {Jim{\'e}nez-Hoyos}, \citenamefont {Sun},\ and\ \citenamefont {Chan}}]{Wouters2016}%
  \BibitemOpen
  \bibfield  {author} {\bibinfo {author} {\bibfnamefont {S.}~\bibnamefont {Wouters}}, \bibinfo {author} {\bibfnamefont {C.~A.}\ \bibnamefont {Jim{\'e}nez-Hoyos}}, \bibinfo {author} {\bibfnamefont {Q.}~\bibnamefont {Sun}},\ and\ \bibinfo {author} {\bibfnamefont {G.~K.-L.}\ \bibnamefont {Chan}},\ }\href@noop {} {\bibfield  {journal} {\bibinfo  {journal} {J. Chem. Theory Comput.}\ }\textbf {\bibinfo {volume} {12}},\ \bibinfo {pages} {2706} (\bibinfo {year} {2016})}\BibitemShut {NoStop}%
\bibitem [{\citenamefont {Cui}\ \emph {et~al.}(2020)\citenamefont {Cui}, \citenamefont {Zhu},\ and\ \citenamefont {Chan}}]{Zhu2020b}%
  \BibitemOpen
  \bibfield  {author} {\bibinfo {author} {\bibfnamefont {Z.-H.}\ \bibnamefont {Cui}}, \bibinfo {author} {\bibfnamefont {T.}~\bibnamefont {Zhu}},\ and\ \bibinfo {author} {\bibfnamefont {G.~K.-L.}\ \bibnamefont {Chan}},\ }\href {https://doi.org/10.1021/acs.jctc.9b00933} {\bibfield  {journal} {\bibinfo  {journal} {J. Chem. Theory Comput.}\ }\textbf {\bibinfo {volume} {16}},\ \bibinfo {pages} {119} (\bibinfo {year} {2020})},\ \bibinfo {note} {pMID: 31815466}\BibitemShut {NoStop}%
\bibitem [{\citenamefont {Sekaran}\ \emph {et~al.}(2021)\citenamefont {Sekaran}, \citenamefont {Tsuchiizu}, \citenamefont {Sauban\`ere},\ and\ \citenamefont {Fromager}}]{Sekaran2021}%
  \BibitemOpen
  \bibfield  {author} {\bibinfo {author} {\bibfnamefont {S.}~\bibnamefont {Sekaran}}, \bibinfo {author} {\bibfnamefont {M.}~\bibnamefont {Tsuchiizu}}, \bibinfo {author} {\bibfnamefont {M.}~\bibnamefont {Sauban\`ere}},\ and\ \bibinfo {author} {\bibfnamefont {E.}~\bibnamefont {Fromager}},\ }\href {https://doi.org/10.1103/PhysRevB.104.035121} {\bibfield  {journal} {\bibinfo  {journal} {Phys. Rev. B}\ }\textbf {\bibinfo {volume} {104}},\ \bibinfo {pages} {035121} (\bibinfo {year} {2021})}\BibitemShut {NoStop}%
\bibitem [{\citenamefont {Sekaran}\ \emph {et~al.}(2023)\citenamefont {Sekaran}, \citenamefont {Bindech},\ and\ \citenamefont {Fromager}}]{Sekaran2023}%
  \BibitemOpen
  \bibfield  {author} {\bibinfo {author} {\bibfnamefont {S.}~\bibnamefont {Sekaran}}, \bibinfo {author} {\bibfnamefont {O.}~\bibnamefont {Bindech}},\ and\ \bibinfo {author} {\bibfnamefont {E.}~\bibnamefont {Fromager}},\ }\href@noop {} {\bibfield  {journal} {\bibinfo  {journal} {J. Chem. Phys.}\ }\textbf {\bibinfo {volume} {159}},\ \bibinfo {pages} {034107} (\bibinfo {year} {2023})}\BibitemShut {NoStop}%
\bibitem [{\citenamefont {Fertitta}\ and\ \citenamefont {Booth}(2018)}]{Fertitta2018}%
  \BibitemOpen
  \bibfield  {author} {\bibinfo {author} {\bibfnamefont {E.}~\bibnamefont {Fertitta}}\ and\ \bibinfo {author} {\bibfnamefont {G.~H.}\ \bibnamefont {Booth}},\ }\href@noop {} {\bibfield  {journal} {\bibinfo  {journal} {Phys. Rev. B}\ }\textbf {\bibinfo {volume} {98}},\ \bibinfo {pages} {235132} (\bibinfo {year} {2018})}\BibitemShut {NoStop}%
\bibitem [{\citenamefont {Sriluckshmy}\ \emph {et~al.}(2021)\citenamefont {Sriluckshmy}, \citenamefont {Nusspickel}, \citenamefont {Fertitta},\ and\ \citenamefont {Booth}}]{Sriluckshmy2021}%
  \BibitemOpen
  \bibfield  {author} {\bibinfo {author} {\bibfnamefont {P.~V.}\ \bibnamefont {Sriluckshmy}}, \bibinfo {author} {\bibfnamefont {M.}~\bibnamefont {Nusspickel}}, \bibinfo {author} {\bibfnamefont {E.}~\bibnamefont {Fertitta}},\ and\ \bibinfo {author} {\bibfnamefont {G.~H.}\ \bibnamefont {Booth}},\ }\href {https://doi.org/10.1103/PhysRevB.103.085131} {\bibfield  {journal} {\bibinfo  {journal} {Phys. Rev. B}\ }\textbf {\bibinfo {volume} {103}},\ \bibinfo {pages} {085131} (\bibinfo {year} {2021})}\BibitemShut {NoStop}%
\bibitem [{\citenamefont {Kananenka}\ \emph {et~al.}(2015)\citenamefont {Kananenka}, \citenamefont {Gull},\ and\ \citenamefont {Zgid}}]{Kananenka2015}%
  \BibitemOpen
  \bibfield  {author} {\bibinfo {author} {\bibfnamefont {A.~A.}\ \bibnamefont {Kananenka}}, \bibinfo {author} {\bibfnamefont {E.}~\bibnamefont {Gull}},\ and\ \bibinfo {author} {\bibfnamefont {D.}~\bibnamefont {Zgid}},\ }\href {https://doi.org/10.1103/PhysRevB.91.121111} {\bibfield  {journal} {\bibinfo  {journal} {Phys. Rev. B}\ }\textbf {\bibinfo {volume} {91}},\ \bibinfo {pages} {121111} (\bibinfo {year} {2015})}\BibitemShut {NoStop}%
\bibitem [{\citenamefont {Lan}\ \emph {et~al.}(2015)\citenamefont {Lan}, \citenamefont {Kananenka},\ and\ \citenamefont {Zgid}}]{Lan2015}%
  \BibitemOpen
  \bibfield  {author} {\bibinfo {author} {\bibfnamefont {T.~N.}\ \bibnamefont {Lan}}, \bibinfo {author} {\bibfnamefont {A.~A.}\ \bibnamefont {Kananenka}},\ and\ \bibinfo {author} {\bibfnamefont {D.}~\bibnamefont {Zgid}},\ }\href@noop {} {\bibfield  {journal} {\bibinfo  {journal} {J. Chem. Phys.}\ }\textbf {\bibinfo {volume} {143}},\ \bibinfo {pages} {241102} (\bibinfo {year} {2015})}\BibitemShut {NoStop}%
\bibitem [{\citenamefont {Iskakov}\ \emph {et~al.}(2020)\citenamefont {Iskakov}, \citenamefont {Yeh}, \citenamefont {Gull},\ and\ \citenamefont {Zgid}}]{Iskakov2020}%
  \BibitemOpen
  \bibfield  {author} {\bibinfo {author} {\bibfnamefont {S.}~\bibnamefont {Iskakov}}, \bibinfo {author} {\bibfnamefont {C.-N.}\ \bibnamefont {Yeh}}, \bibinfo {author} {\bibfnamefont {E.}~\bibnamefont {Gull}},\ and\ \bibinfo {author} {\bibfnamefont {D.}~\bibnamefont {Zgid}},\ }\href {https://doi.org/10.1103/PhysRevB.102.085105} {\bibfield  {journal} {\bibinfo  {journal} {Phys. Rev. B}\ }\textbf {\bibinfo {volume} {102}},\ \bibinfo {pages} {085105} (\bibinfo {year} {2020})}\BibitemShut {NoStop}%
\bibitem [{\citenamefont {Mejuto-Zaera}(2024)}]{Mejuto2024}%
  \BibitemOpen
  \bibfield  {author} {\bibinfo {author} {\bibfnamefont {C.}~\bibnamefont {Mejuto-Zaera}},\ }\href {https://doi.org/10.1039/D4FD00053F} {\bibfield  {journal} {\bibinfo  {journal} {Faraday Discuss.}\ }\textbf {\bibinfo {volume} {254}},\ \bibinfo {pages} {653} (\bibinfo {year} {2024})}\BibitemShut {NoStop}%
\bibitem [{\citenamefont {Pasqua}\ \emph {et~al.}(2025)\citenamefont {Pasqua}, \citenamefont {Tagliente}, \citenamefont {Bellomia}, \citenamefont {Monserrat}, \citenamefont {Fabrizio},\ and\ \citenamefont {Mejuto-Zaera}}]{Pasqua2025}%
  \BibitemOpen
  \bibfield  {author} {\bibinfo {author} {\bibfnamefont {I.}~\bibnamefont {Pasqua}}, \bibinfo {author} {\bibfnamefont {A.~M.}\ \bibnamefont {Tagliente}}, \bibinfo {author} {\bibfnamefont {G.}~\bibnamefont {Bellomia}}, \bibinfo {author} {\bibfnamefont {B.}~\bibnamefont {Monserrat}}, \bibinfo {author} {\bibfnamefont {M.}~\bibnamefont {Fabrizio}},\ and\ \bibinfo {author} {\bibfnamefont {C.}~\bibnamefont {Mejuto-Zaera}},\ }\href {https://arxiv.org/abs/2507.10670} {\bibinfo {title} {Quasiparticle band picture bridging topology and strong correlations across energy scales}} (\bibinfo {year} {2025}),\ \Eprint {https://arxiv.org/abs/2507.10670} {arXiv:2507.10670 [cond-mat.str-el]} \BibitemShut {NoStop}%
\bibitem [{\citenamefont {Giuli}\ \emph {et~al.}(2025)\citenamefont {Giuli}, \citenamefont {Mejuto-Zaera},\ and\ \citenamefont {Capone}}]{Giuli2025}%
  \BibitemOpen
  \bibfield  {author} {\bibinfo {author} {\bibfnamefont {S.}~\bibnamefont {Giuli}}, \bibinfo {author} {\bibfnamefont {C.}~\bibnamefont {Mejuto-Zaera}},\ and\ \bibinfo {author} {\bibfnamefont {M.}~\bibnamefont {Capone}},\ }\href {https://doi.org/10.1103/PhysRevB.111.L020401} {\bibfield  {journal} {\bibinfo  {journal} {Phys. Rev. B}\ }\textbf {\bibinfo {volume} {111}},\ \bibinfo {pages} {L020401} (\bibinfo {year} {2025})}\BibitemShut {NoStop}%
\bibitem [{\citenamefont {Bellomia}\ \emph {et~al.}(2025)\citenamefont {Bellomia}, \citenamefont {Amaricci},\ and\ \citenamefont {Capone}}]{Bellomia_intracorr}%
  \BibitemOpen
  \bibfield  {author} {\bibinfo {author} {\bibfnamefont {G.}~\bibnamefont {Bellomia}}, \bibinfo {author} {\bibfnamefont {A.}~\bibnamefont {Amaricci}},\ and\ \bibinfo {author} {\bibfnamefont {M.}~\bibnamefont {Capone}},\ }\href {https://arxiv.org/abs/2506.18709} {\bibinfo {title} {Local classical correlations between physical electrons in the hubbard model}} (\bibinfo {year} {2025}),\ \Eprint {https://arxiv.org/abs/2506.18709} {arXiv:2506.18709 [cond-mat.str-el]} \BibitemShut {NoStop}%
\bibitem [{\citenamefont {Tagliente}\ \emph {et~al.}(2025)\citenamefont {Tagliente}, \citenamefont {Mejuto-Zaera},\ and\ \citenamefont {Fabrizio}}]{Tagliente2025}%
  \BibitemOpen
  \bibfield  {author} {\bibinfo {author} {\bibfnamefont {A.~M.}\ \bibnamefont {Tagliente}}, \bibinfo {author} {\bibfnamefont {C.}~\bibnamefont {Mejuto-Zaera}},\ and\ \bibinfo {author} {\bibfnamefont {M.}~\bibnamefont {Fabrizio}},\ }\href {https://doi.org/10.1103/PhysRevB.111.125110} {\bibfield  {journal} {\bibinfo  {journal} {Phys. Rev. B}\ }\textbf {\bibinfo {volume} {111}},\ \bibinfo {pages} {125110} (\bibinfo {year} {2025})}\BibitemShut {NoStop}%
\bibitem [{\citenamefont {Yao}\ \emph {et~al.}(2021)\citenamefont {Yao}, \citenamefont {Zhang}, \citenamefont {Wang}, \citenamefont {Ho},\ and\ \citenamefont {Orth}}]{Yao2021}%
  \BibitemOpen
  \bibfield  {author} {\bibinfo {author} {\bibfnamefont {Y.}~\bibnamefont {Yao}}, \bibinfo {author} {\bibfnamefont {F.}~\bibnamefont {Zhang}}, \bibinfo {author} {\bibfnamefont {C.-Z.}\ \bibnamefont {Wang}}, \bibinfo {author} {\bibfnamefont {K.-M.}\ \bibnamefont {Ho}},\ and\ \bibinfo {author} {\bibfnamefont {P.~P.}\ \bibnamefont {Orth}},\ }\href {https://doi.org/10.1103/PhysRevResearch.3.013184} {\bibfield  {journal} {\bibinfo  {journal} {Phys. Rev. Res.}\ }\textbf {\bibinfo {volume} {3}},\ \bibinfo {pages} {013184} (\bibinfo {year} {2021})}\BibitemShut {NoStop}%
\bibitem [{\citenamefont {Besserve}\ and\ \citenamefont {Ayral}(2022)}]{Besserve2022}%
  \BibitemOpen
  \bibfield  {author} {\bibinfo {author} {\bibfnamefont {P.}~\bibnamefont {Besserve}}\ and\ \bibinfo {author} {\bibfnamefont {T.}~\bibnamefont {Ayral}},\ }\href {https://doi.org/10.1103/PhysRevB.105.115108} {\bibfield  {journal} {\bibinfo  {journal} {Phys. Rev. B}\ }\textbf {\bibinfo {volume} {105}},\ \bibinfo {pages} {115108} (\bibinfo {year} {2022})}\BibitemShut {NoStop}%
\bibitem [{\citenamefont {Chen}\ \emph {et~al.}(2025)\citenamefont {Chen}, \citenamefont {Khindanov}, \citenamefont {Salazar}, \citenamefont {Barona}, \citenamefont {Zhang}, \citenamefont {Wang}, \citenamefont {Iadecola}, \citenamefont {Lanatà},\ and\ \citenamefont {Yao}}]{Chen2025}%
  \BibitemOpen
  \bibfield  {author} {\bibinfo {author} {\bibfnamefont {I.-C.}\ \bibnamefont {Chen}}, \bibinfo {author} {\bibfnamefont {A.}~\bibnamefont {Khindanov}}, \bibinfo {author} {\bibfnamefont {C.}~\bibnamefont {Salazar}}, \bibinfo {author} {\bibfnamefont {H.~M.}\ \bibnamefont {Barona}}, \bibinfo {author} {\bibfnamefont {F.}~\bibnamefont {Zhang}}, \bibinfo {author} {\bibfnamefont {C.-Z.}\ \bibnamefont {Wang}}, \bibinfo {author} {\bibfnamefont {T.}~\bibnamefont {Iadecola}}, \bibinfo {author} {\bibfnamefont {N.}~\bibnamefont {Lanatà}},\ and\ \bibinfo {author} {\bibfnamefont {Y.-X.}\ \bibnamefont {Yao}},\ }\href {https://arxiv.org/abs/2506.01204} {\bibinfo {title} {Quantum-classical embedding via ghost gutzwiller approximation for enhanced simulations of correlated electron systems}} (\bibinfo {year} {2025}),\ \Eprint {https://arxiv.org/abs/2506.01204} {arXiv:2506.01204 [quant-ph]} \BibitemShut {NoStop}%
\bibitem [{\citenamefont {Sriluckshmy}\ \emph {et~al.}(2025)\citenamefont {Sriluckshmy}, \citenamefont {Jamet},\ and\ \citenamefont {Šimkovic IV}}]{Sriluckshmy2025}%
  \BibitemOpen
  \bibfield  {author} {\bibinfo {author} {\bibfnamefont {P.~V.}\ \bibnamefont {Sriluckshmy}}, \bibinfo {author} {\bibfnamefont {F.}~\bibnamefont {Jamet}},\ and\ \bibinfo {author} {\bibfnamefont {F.}~\bibnamefont {Šimkovic IV}},\ }\href {https://arxiv.org/abs/2506.21431} {\bibinfo {title} {Quantum assisted ghost gutzwiller ansatz}} (\bibinfo {year} {2025}),\ \Eprint {https://arxiv.org/abs/2506.21431} {arXiv:2506.21431 [quant-ph]} \BibitemShut {NoStop}%
\bibitem [{\citenamefont {Sun}\ \emph {et~al.}(2020)\citenamefont {Sun}, \citenamefont {Zhang}, \citenamefont {Banerjee}, \citenamefont {Bao}, \citenamefont {Barbry}, \citenamefont {Blunt}, \citenamefont {Bogdanov}, \citenamefont {Booth}, \citenamefont {Chen}, \citenamefont {Cui}, \citenamefont {Eriksen}, \citenamefont {Gao}, \citenamefont {Guo}, \citenamefont {Hermann}, \citenamefont {Hermes}, \citenamefont {Koh}, \citenamefont {Koval}, \citenamefont {Lehtola}, \citenamefont {Li}, \citenamefont {Liu}, \citenamefont {Mardirossian}, \citenamefont {McClain}, \citenamefont {Motta}, \citenamefont {Mussard}, \citenamefont {Pham}, \citenamefont {Pulkin}, \citenamefont {Purwanto}, \citenamefont {Robinson}, \citenamefont {Ronca}, \citenamefont {Sayfutyarova}, \citenamefont {Scheurer}, \citenamefont {Schurkus}, \citenamefont {Smith}, \citenamefont {Sun}, \citenamefont {Sun}, \citenamefont {Upadhyay}, \citenamefont {Wagner}, \citenamefont {Wang}, \citenamefont {White}, \citenamefont {Whitfield}, \citenamefont
  {Williamson}, \citenamefont {Wouters}, \citenamefont {Yang}, \citenamefont {Yu}, \citenamefont {Zhu}, \citenamefont {Berkelbach}, \citenamefont {Sharma}, \citenamefont {Sokolov},\ and\ \citenamefont {Chan}}]{PYSCF1}%
  \BibitemOpen
  \bibfield  {author} {\bibinfo {author} {\bibfnamefont {Q.}~\bibnamefont {Sun}}, \bibinfo {author} {\bibfnamefont {X.}~\bibnamefont {Zhang}}, \bibinfo {author} {\bibfnamefont {S.}~\bibnamefont {Banerjee}}, \bibinfo {author} {\bibfnamefont {P.}~\bibnamefont {Bao}}, \bibinfo {author} {\bibfnamefont {M.}~\bibnamefont {Barbry}}, \bibinfo {author} {\bibfnamefont {N.~S.}\ \bibnamefont {Blunt}}, \bibinfo {author} {\bibfnamefont {N.~A.}\ \bibnamefont {Bogdanov}}, \bibinfo {author} {\bibfnamefont {G.~H.}\ \bibnamefont {Booth}}, \bibinfo {author} {\bibfnamefont {J.}~\bibnamefont {Chen}}, \bibinfo {author} {\bibfnamefont {Z.-H.}\ \bibnamefont {Cui}}, \bibinfo {author} {\bibfnamefont {J.~J.}\ \bibnamefont {Eriksen}}, \bibinfo {author} {\bibfnamefont {Y.}~\bibnamefont {Gao}}, \bibinfo {author} {\bibfnamefont {S.}~\bibnamefont {Guo}}, \bibinfo {author} {\bibfnamefont {J.}~\bibnamefont {Hermann}}, \bibinfo {author} {\bibfnamefont {M.~R.}\ \bibnamefont {Hermes}}, \bibinfo {author} {\bibfnamefont {K.}~\bibnamefont {Koh}},
  \bibinfo {author} {\bibfnamefont {P.}~\bibnamefont {Koval}}, \bibinfo {author} {\bibfnamefont {S.}~\bibnamefont {Lehtola}}, \bibinfo {author} {\bibfnamefont {Z.}~\bibnamefont {Li}}, \bibinfo {author} {\bibfnamefont {J.}~\bibnamefont {Liu}}, \bibinfo {author} {\bibfnamefont {N.}~\bibnamefont {Mardirossian}}, \bibinfo {author} {\bibfnamefont {J.~D.}\ \bibnamefont {McClain}}, \bibinfo {author} {\bibfnamefont {M.}~\bibnamefont {Motta}}, \bibinfo {author} {\bibfnamefont {B.}~\bibnamefont {Mussard}}, \bibinfo {author} {\bibfnamefont {H.~Q.}\ \bibnamefont {Pham}}, \bibinfo {author} {\bibfnamefont {A.}~\bibnamefont {Pulkin}}, \bibinfo {author} {\bibfnamefont {W.}~\bibnamefont {Purwanto}}, \bibinfo {author} {\bibfnamefont {P.~J.}\ \bibnamefont {Robinson}}, \bibinfo {author} {\bibfnamefont {E.}~\bibnamefont {Ronca}}, \bibinfo {author} {\bibfnamefont {E.~R.}\ \bibnamefont {Sayfutyarova}}, \bibinfo {author} {\bibfnamefont {M.}~\bibnamefont {Scheurer}}, \bibinfo {author} {\bibfnamefont {H.~F.}\ \bibnamefont {Schurkus}},
  \bibinfo {author} {\bibfnamefont {J.~E.~T.}\ \bibnamefont {Smith}}, \bibinfo {author} {\bibfnamefont {C.}~\bibnamefont {Sun}}, \bibinfo {author} {\bibfnamefont {S.-N.}\ \bibnamefont {Sun}}, \bibinfo {author} {\bibfnamefont {S.}~\bibnamefont {Upadhyay}}, \bibinfo {author} {\bibfnamefont {L.~K.}\ \bibnamefont {Wagner}}, \bibinfo {author} {\bibfnamefont {X.}~\bibnamefont {Wang}}, \bibinfo {author} {\bibfnamefont {A.}~\bibnamefont {White}}, \bibinfo {author} {\bibfnamefont {J.~D.}\ \bibnamefont {Whitfield}}, \bibinfo {author} {\bibfnamefont {M.~J.}\ \bibnamefont {Williamson}}, \bibinfo {author} {\bibfnamefont {S.}~\bibnamefont {Wouters}}, \bibinfo {author} {\bibfnamefont {J.}~\bibnamefont {Yang}}, \bibinfo {author} {\bibfnamefont {J.~M.}\ \bibnamefont {Yu}}, \bibinfo {author} {\bibfnamefont {T.}~\bibnamefont {Zhu}}, \bibinfo {author} {\bibfnamefont {T.~C.}\ \bibnamefont {Berkelbach}}, \bibinfo {author} {\bibfnamefont {S.}~\bibnamefont {Sharma}}, \bibinfo {author} {\bibfnamefont {A.~Y.}\ \bibnamefont
  {Sokolov}},\ and\ \bibinfo {author} {\bibfnamefont {G.~K.-L.}\ \bibnamefont {Chan}},\ }\href {https://doi.org/10.1063/5.0006074} {\bibfield  {journal} {\bibinfo  {journal} {The Journal of Chemical Physics}\ }\textbf {\bibinfo {volume} {153}},\ \bibinfo {pages} {024109} (\bibinfo {year} {2020})}\BibitemShut {NoStop}%
\bibitem [{\citenamefont {Sun}\ \emph {et~al.}(2018)\citenamefont {Sun}, \citenamefont {Berkelbach}, \citenamefont {Blunt}, \citenamefont {Booth}, \citenamefont {Guo}, \citenamefont {Li}, \citenamefont {Liu}, \citenamefont {McClain}, \citenamefont {Sayfutyarova}, \citenamefont {Sharma}, \citenamefont {Wouters},\ and\ \citenamefont {Chan}}]{PYSCF2}%
  \BibitemOpen
  \bibfield  {author} {\bibinfo {author} {\bibfnamefont {Q.}~\bibnamefont {Sun}}, \bibinfo {author} {\bibfnamefont {T.~C.}\ \bibnamefont {Berkelbach}}, \bibinfo {author} {\bibfnamefont {N.~S.}\ \bibnamefont {Blunt}}, \bibinfo {author} {\bibfnamefont {G.~H.}\ \bibnamefont {Booth}}, \bibinfo {author} {\bibfnamefont {S.}~\bibnamefont {Guo}}, \bibinfo {author} {\bibfnamefont {Z.}~\bibnamefont {Li}}, \bibinfo {author} {\bibfnamefont {J.}~\bibnamefont {Liu}}, \bibinfo {author} {\bibfnamefont {J.~D.}\ \bibnamefont {McClain}}, \bibinfo {author} {\bibfnamefont {E.~R.}\ \bibnamefont {Sayfutyarova}}, \bibinfo {author} {\bibfnamefont {S.}~\bibnamefont {Sharma}}, \bibinfo {author} {\bibfnamefont {S.}~\bibnamefont {Wouters}},\ and\ \bibinfo {author} {\bibfnamefont {G.~K.-L.}\ \bibnamefont {Chan}},\ }\href@noop {} {\bibfield  {journal} {\bibinfo  {journal} {WIREs Computational Molecular Science}\ }\textbf {\bibinfo {volume} {8}},\ \bibinfo {pages} {e1340} (\bibinfo {year} {2018})}\BibitemShut {NoStop}%
\bibitem [{\citenamefont {Sun}(2015)}]{PYSCF3}%
  \BibitemOpen
  \bibfield  {author} {\bibinfo {author} {\bibfnamefont {Q.}~\bibnamefont {Sun}},\ }\href {https://doi.org/https://doi.org/10.1002/jcc.23981} {\bibfield  {journal} {\bibinfo  {journal} {Journal of Computational Chemistry}\ }\textbf {\bibinfo {volume} {36}},\ \bibinfo {pages} {1664} (\bibinfo {year} {2015})},\ \Eprint {https://arxiv.org/abs/https://onlinelibrary.wiley.com/doi/pdf/10.1002/jcc.23981} {https://onlinelibrary.wiley.com/doi/pdf/10.1002/jcc.23981} \BibitemShut {NoStop}%
\bibitem [{\citenamefont {Mejuto-Zaera}\ \emph {et~al.}(2020)\citenamefont {Mejuto-Zaera}, \citenamefont {Zepeda-N{\'u}{\~n}ez}, \citenamefont {Lindsey}, \citenamefont {Tubman}, \citenamefont {Whaley},\ and\ \citenamefont {Lin}}]{Mejuto2020}%
  \BibitemOpen
  \bibfield  {author} {\bibinfo {author} {\bibfnamefont {C.}~\bibnamefont {Mejuto-Zaera}}, \bibinfo {author} {\bibfnamefont {L.}~\bibnamefont {Zepeda-N{\'u}{\~n}ez}}, \bibinfo {author} {\bibfnamefont {M.}~\bibnamefont {Lindsey}}, \bibinfo {author} {\bibfnamefont {N.}~\bibnamefont {Tubman}}, \bibinfo {author} {\bibfnamefont {B.}~\bibnamefont {Whaley}},\ and\ \bibinfo {author} {\bibfnamefont {L.}~\bibnamefont {Lin}},\ }\href@noop {} {\bibfield  {journal} {\bibinfo  {journal} {Physical Review B}\ }\textbf {\bibinfo {volume} {101}},\ \bibinfo {pages} {035143} (\bibinfo {year} {2020})}\BibitemShut {NoStop}%
\bibitem [{\citenamefont {Florez-Ablan}\ \emph {et~al.}(2025)\citenamefont {Florez-Ablan}, \citenamefont {Mejuto-Zaera},\ and\ \citenamefont {Capone}}]{Florez2025}%
  \BibitemOpen
  \bibfield  {author} {\bibinfo {author} {\bibfnamefont {D.}~\bibnamefont {Florez-Ablan}}, \bibinfo {author} {\bibfnamefont {C.}~\bibnamefont {Mejuto-Zaera}},\ and\ \bibinfo {author} {\bibfnamefont {M.}~\bibnamefont {Capone}},\ }\href {https://doi.org/10.1103/lnhv-kvy4} {\bibfield  {journal} {\bibinfo  {journal} {Phys. Rev. B}\ }\textbf {\bibinfo {volume} {112}},\ \bibinfo {pages} {085153} (\bibinfo {year} {2025})}\BibitemShut {NoStop}%
\end{thebibliography}

\end{document}